\documentclass[12pt,preprint]{aastex}
\usepackage{times}
\usepackage{enumerate}
\usepackage{graphicx}
\usepackage{natbib}
\usepackage{epstopdf}
\usepackage{rotating}
\newcommand{\ltsima}{\stackrel{\textstyle <}{\sim}}
\newcommand{\simlt}{\scriptsize{\raisebox{-2pt}{$\ltsima$}}\normalsize}
\newcommand{\gtsima}{\stackrel{\textstyle >}{\sim}}
\newcommand{\simgt}{\scriptsize{\raisebox{-2pt}{$\gtsima$}}\normalsize}

\renewcommand{\thepage}{\rm\arabic{page}}

\newcommand{\water}{H$_2$O}

\newcommand{\oxy}{O$_2$}
\newcommand{\noxy}{$N$(O$_2$)}
\newcommand{\aoxy}{$\chi$(O$_2$)}
\newcommand{\nco}{$^{12}$CO}
\newcommand{\ico}{$^{13}$CO}
\newcommand{\cio}{C$^{18}$O}

\newcommand{\asec}{$^{\prime\prime}$}
\newcommand{\amin}{$^{\prime}$}

\newcommand{\mh}{H$_2$}
\newcommand{\nh}{$n${(H$_2$)}}

\newcommand{\av}{{\em A}$_{\rm V}$}

\newcommand{\kms}{~km~s$^{{-1}}$}
\newcommand{\cmc}{cm$^{-3}$}
\newcommand{\cms}{cm$^{-2}$}

\newcommand{\go}{$G_{\rm o}$}
\newcommand{\dash}{$\,$--$\,$}

\newcommand{\etal}{et$\,$al.}
\newcommand{\ti}{$\,\times\,$}

\renewcommand{\apj}{{ApJ}}
\renewcommand{\apjl}{{ApJL}}
\renewcommand{\apjs}{{ApJS}}

\renewcommand{\mnras}{{MNRAS}}

\renewcommand{\aap}{{A\&A}}

\makeatletter
\def\fnum@figure{{Fig.~\thefigure}}
\long\def\@makecaption#1#2{
 \vskip 10pt 
 \setbox\@tempboxa\hbox{#1. #2}
 \ifdim \wd\@tempboxa >\hsize \unhbox\@tempboxa\par \else \hbox
to\hsize{\hfil\box\@tempboxa\hfil} 
 \fi}

\makeatother

\makeatletter
\def\ps@myheadings{\let\@mkboth\@gobbletwo
\def\@oddhead{\hbox{}\sl\rightmark \hfil \thepage}%
\def\@oddfoot{}\def\@evenhead{\thepage\hfil\sl\leftmark\hbox {}}%
\def\@evenfoot{}\def\sectionmark##1{}\def\subsectionmark##1{}}
\makeatother

\makeatletter
\def\subsection{\@startsection{subsection}{2}{\z@}{-3.25ex plus -1ex minus 
   -.2ex}{1.5ex plus .2ex}{ \em \rm}}
\makeatother

\newcounter{ncount}

\shorttitle{The Effects of FUV and X-Ray Illumination on Magnetohydrodynamic Shock Waves}
\shortauthors{Kaufman and Melnick}

\begin{document}

\begin{center}
{\large\bf O$_2$ Emission Toward Orion H$_2$ Peak 1}\\*[1.5mm]
{\large\bf and the Role of FUV-Illuminated C-Shocks} \\*[9mm]
Gary Melnick$^1$ and Michael J. Kaufman$^2$, \\*[1.4in]
{\large\em To appear in the Astrophysical Journal}

\vspace{24mm}


\vspace{3.5in}

Received$\;$\rule{1.65in}{0.25mm}\hspace{-1.3in}\raisebox{1mm}{Feb. 19, 2015}\hspace{0.46in};~~~~~Accepted$\;$\rule{1.65in}{0.25mm}\hspace{-1.3in}\raisebox{1mm}{May 4, 2015}\end{center}


\clearpage

\setcounter{page}{2}

\pagestyle{myheadings}
\markright{{}\hfill{\hbox{\hss\rm Page }}}

\begin{enumerate}
\item Harvard-Smithsonian Center for Astrophysics, 60 Garden Street, MS 66, Cambridge, MA 02138, USA \\*[-6mm]

\item Department of Physics and Astronomy, San Jos\'{e} State University, San Jos\'{e}, CA 95192, USA \\*[-6mm]

\end{enumerate}


\renewcommand{\baselinestretch}{1.27}

\begin{abstract}
\vspace{-5mm}
Molecular oxygen (\oxy) has been the target of ground-based and space-borne 
searches for decades.  Of the thousands of lines of sight surveyed, only those
toward Rho Ophiuchus and Orion \mh\ Peak 1 have yielded detections of any statistical significance.
The detection of the \oxy\ $N_J =\,$3$_3$\dash 1$_2$ and 5$_4$\dash 3$_4$ lines at 487.249 GHz and 773.840 GHz, respectively, toward Rho Ophiuchus has been attributed to a short-lived peak in the time-dependent,
cold-cloud \oxy\ abundance, while the detection of the \oxy\ $N_J =\,$3$_3$\dash 1$_2$, 5$_4$\dash 3$_4$ lines, 
plus the 7$_6$\dash 5$_6$ line at 1120.715 GHz, toward Orion has been ascribed to time-dependent
preshock physical and chemical evolution and low-velocity (12\kms) non-dissociative $C$-type shocks, both of which
are fully shielded from far-ultraviolet (FUV) radiation, plus a postshock region that is exposed to a FUV field.
We report a re-interpretation of the Orion \oxy\ detection
based on new $C$-type shock models that fully incorporate the significant effects the presence of even a weak
FUV field can have on the preshock gas, shock structure and postshock chemistry.  In particular, we show that a family
of solutions exists,  depending on the FUV intensity, that reproduces {\em both} the observed \oxy\ intensities and 
\oxy\ line ratios.
The solution in closest agreement with the shock parameters inferred for \mh\ Peak 1 from other gas tracers assumes
a 23\kms\ shock impacting gas with a preshock density of 8\ti 10$^4$~\cmc\ and \go$=\,$1, substantially different 
from that inferred for the fully-shielded shock case.  As pointed out previously, the similarity between the LSR 
velocity of all three \oxy\ lines ($\approx\,$11\kms) and recently measured \water\ 5$_{32}$\dash 4$_{41}$ 
maser emission at 620.701 GHz toward \mh\ 
Peak 1 suggests that the \oxy\ emission arises behind the same shocks responsible for the maser emission, though
the \oxy\ emission is almost certainly more extended than the localized high-density maser spots.  Since
maser emission arises along  lines of sight of low-velocity gradient, indicating shock motion largely perpendicular to our 
line of sight,  we note that this geometry can not only explain the narrow 
($\simlt\,$3\kms) observed \oxy\ line widths despite their excitation behind a shock, but also why such \oxy\
detections are rare.
\end{abstract}


\renewcommand{\baselinestretch}{1.0}

\vspace{4.5in}

\baselineskip=24pt

\pagebreak

\section{INTRODUCTION}

Accounting for the low abundance of molecular oxygen (\oxy) in dense (\nh$\;\simgt\,$10$^3\;$\cmc) 
molecular clouds has long posed a challenge to both observers and theorists.  For almost 30 years,
the search for \oxy\ was motivated by predictions of gas-phase chemical models that \oxy\ was a
major reservoir of elemental oxygen within dense clouds \citep[e.g.][]{Goldsmith78,Neufeld95}.  
However, the inability of gas-phase models to account for the observed abundance of an increasing number of species
highlighted the shortcomings of such models.  This became
evident again following the launch of the {\em Submillimeter Wave Astronomy
Satellite (SWAS)} \citep{Melnick00} and {\em Odin} \citep{Nordh03} when their largely unsuccessful searches
for \oxy\ established upper limits to the \oxy\ abundance more than 100 times below the predictions of these models
\citep[e.g.,][]{Goldsmith00}.
Modified chemical models, which include the effects of dust grains as sites for molecule formation as well as 
freeze out, were then invoked to explain the low \oxy\ abundance and weak emission
\citep[cf.][]{Bergin00, Hollenbach09}.  With the launch of the {\em Herschel Space Observatory}
\citep{Pilbratt10}, attention turned to using its greater sensitivity to target regions predicted by these
updated chemical models to possess high columns of warm \oxy.

The Open Time Key Program, {\em Herschel} Oxygen Project (HOP), was proposed and selected to carry out a 
survey of warm molecular clouds in the following rotational transitions of \oxy:~({\em N,$\;$J})$\,=\,$(3,3) 
$\rightarrow\,$(1,2) at 487 GHz, (5,4) $\rightarrow\,$(3,4) at 774 GHz, and (7,6) $\rightarrow\,$(5,6) at 1121 GHz
\citep[see][]{Goldsmith11}.  Among the sources
of particular interest was the Orion Molecular Cloud because of the large column densities of warm 
molecular gas known to exist toward several
prominent components, including the Hot Core, the Orion Bar, and \mh\ Peaks 1 and 2.
While no convincing \oxy\ emission was detected toward the Hot Core \citep{Goldsmith11} or
Orion Bar \citep{Melnick12}, \oxy\ emission has been detected toward \mh\ Peak 1 
\citep{Goldsmith11, Chen14}.  This detection raises two questions: (1) What plausible conditions
exist at \mh\ Peak 1 that could produce the observed \oxy\ line intensities {\em and} line ratios?; and, (2)
Are these conditions nonetheless sufficiently rare to explain the absence of 
detectable \oxy\ emission toward most other sources searched?

Intense \mh\ emission toward Peak 1 is observed in transitions with excitation energies, $E_u/k$,
ranging from 1015 to 43,000~K that
has long been attributed to collisional excitations due to shock waves presumed to result from the high-velocity 
outflow originating $\sim\,$30\asec\ to the SE in the BN/KL region \citep{Gautier76, Beckwith78, Rosenthal00}.  
The Orion Molecular Cloud is
also subject to strong far-ultraviolet (FUV) radiation from sources such as $\theta^1$C~Ori \cite[cf.][]{Kristensen03}.
In this paper, we examine the role FUV radiation plays in altering the chemical and physical
conditions in gas capable of reproducing the \oxy\ observations reported in \citet{Chen14}.  In particular,
we focus on FUV-illuminated gas subject to the passage of non-dissociative $C$-type shocks as regions
especially conducive to the production of elevated \oxy\ abundances.

Fully shielded non-dissociative shocks are generally inefficient at producing \oxy\ in the postshocked gas.
Even when shock velocities are sufficient to sputter material from grain surfaces, all but $\sim\,$1\% of the
gas-phase oxygen not locked in CO is rapidly processed into \water\ within the warm ($T~\simgt\:$400$\:$K)
postshock gas via a set of neutral-neutral chemical reactions 
(O$+$\mh$\,\rightarrow\,$OH$+$H; OH$+$\mh$\,\rightarrow\,$\water$+$H).  In the absence of a FUV field,
the \water\ abundance is predicted to remain high throughout the postshock
region and the \oxy\ abundance never exceeds a few percent of that of \water\ \citep[cf.][]{Draine83b, Kaufman96b}.
However, the presence of a FUV field can significantly increase the postshock \oxy\ abundance by: (1) 
increasing the preshock atomic oxygen abundance through both the photodissociation of gas-phase O-bearing 
species and the photodesorption (from grains) and subsequent photodissociation of desorbed O-bearing species, 
leading to a higher postshock \water\ abundance; and, (2) increasing the postshock
abundance of OH and O through the photodissociation of postshock \water, enabling increased production of
\oxy\ via the reaction O$+$OH$\,\rightarrow\,$\oxy$+$H.

In \citet{Chen14}, the three components necessary to model the observed \oxy\ emission -- i.e., the preshock gas,
the shock and its postshock thermal profile, and the postshock chemistry --
were addressed separately.  In particular, the preshock gas and shock were assumed to be fully 
shielded, i.e., no FUV radiation reaches these regions, while the postshock chemistry was modeled 
assuming various levels of FUV illumination.  It is unlikely that the preshock, shock, and postshock gas,
which are in close physical proximity, would experience significantly different FUV fluxes; however, the absence of
a shock code that incorporates the full effects of a FUV field dictated the approach described in Chen et\,al.

In this paper, we re-examine the results presented in Chen et\,al., with several inportant differences.
The main contribution of the present work is the self-consistent treatment of the preshock, shock, and postshock
regions under the influence of FUV fields common to all components.  FUV radiation can not only increase
the preshock atomic oxygen abundance, as described above, but the effects of FUV radiation on the structure 
of $C$-type shocks can be significant.   These effects include
increased postshock temperatures at a given shock velocity, reduced velocities at which these shocks break down 
(and \mh\ is dissociated), as well as the altered abundance of key postshock molecular species.  
We will show that:
(1) the thermal and chemical changes induced by the FUV radiation on the preshock, shock, and postshock gas, and
including the sputtering of \water\ from grain mantles at shock velocities greater than $\sim\,$20\kms, yield multiple
combinations of shock velocity and FUV field intensity capable of producing detectable \oxy\ emission; and, (2) 
the \oxy\ line intensity ratios can be an important discriminator between shock models.  Use of the observed \oxy\ line
ratios was not used as a test of the shock models presented in Chen et\,al.

In addition, the inclusion of a FUV field in all gas components mitigates a problem identified in Chen et\,al.,
i.e., the  need  in their model to assume $\simgt\,$10\% sputtering efficiency of \water\ from grain mantles behind 
a 12\kms\ $C$-type shock.  As will be discussed, a sputtering efficiency of 10\% or
greater behind a 12\kms\ shock is higher than theoretical predictions as well as recent observations.

Finally, the association between co-located \water\ maser emission and the 
\oxy\ emission noted in Chen et\,al.~provides important clues regarding the orientation of the \water- and 
likely the \oxy-producing shocks.
We point out that this orientation not only justifies the assumption of relatively large column densities of \oxy\
along the line of sight, as discussed in Chen et\,al., but also provides
an alternate explanation to that offerred in Chen et\,al. for why such \oxy\ detections are rare.

In \S2, we briefly review the \oxy\ observations upon which our analysis is based.  In \S3, we describe
the modifications to the $C$-type shock models required when such shocks are exposed to FUV radiation.
In \S4, we describe the assumed shock geometry, the methods used to compute the \oxy\ line
integrated intensities, and how well the results fit the observations.  In \S5, we discuss the ways in which
FUV-illuminated $C$-type shocks and the shock geometry toward Orion \mh\ Peak~1 combine to
provide an explanation for the detected \oxy\ emission while also accounting for the apparent rarity
of such emission.

\vspace{1mm}

\section{OBSERVATIONS AND RESULTS}

The {\em Herschel} observational results considered in this paper are presented in \citet{Chen14}.  The interested reader is referred to Chen et\,al.~for details regarding the observations and data reduction; only the
results we model are summarized here.

The observations were conducted in 2012 using the Heterodyne Instrument for 
the Far Infrared \citep[HIFI;][]{deGraauw10} and were centered at J2000 coordinates 
R.A.$\,=\:$5$^{\rm h}$\,35$^{\rm m}$\,14.$\!^{\rm s}$2
and Dec.$\,= -$5$^{\rm o}$\,22$^{\prime}$\,31$^{\prime\prime}$.  
HIFI was used in dual beam switch (DBS) mode with the reference positions located 3\amin\ on either side of 
the source. For each transition, eight local oscillator (LO) settings were used to allow sideband 
deconvolution.  The integration time for each LO setting was 824 seconds for the
487 GHz and 774 GHz spectra, and 3477 seconds for the 1121 GHz spectrum.
The observational results relevant to our modeling effort are presented in Table~1.

\vspace{1mm}

\section{MODIFIED SHOCK MODELS}

\subsection{\em FUV-Illuminated C-Type Shocks}

\vspace{-1mm}

Within interstellar shocks possessing low fractional ionization and a strong magnetic field, the flow variables 
may remain continuous; i.e., the neutral and ionized gases do not experience discontinuities and the 
gas remains relatively cool and molecular. The physics of $C$-type shocks in well-shielded gas has been studied 
extensively \citep[e.g.,][]{Draine83a, Draine83b, Kaufman96a, Flower10}; however, it is increasingly 
evident that FUV radiation from nearby stars plays a role in the physics of molecular outflows 
\citep[e.g.,][]{Kristensen12, vanKempen10}.
In order to assess the role of FUV radiation, we have made several modifications to the \citet{Kaufman96a} model. These modifications affect the abundances of important species in the preshock gas, the shock length scale, and the abundances of oxygen-bearing species in the postshock gas. We briefly describe the effects below; a more detailed description will be included in a forthcoming paper.

\vspace{1mm}

\subsection{\em Preshock Conditions}

\vspace{-1mm}

\begin{figure}[h]
\centering
\vspace{7mm}
$\!\!\!$\includegraphics[scale=0.86]{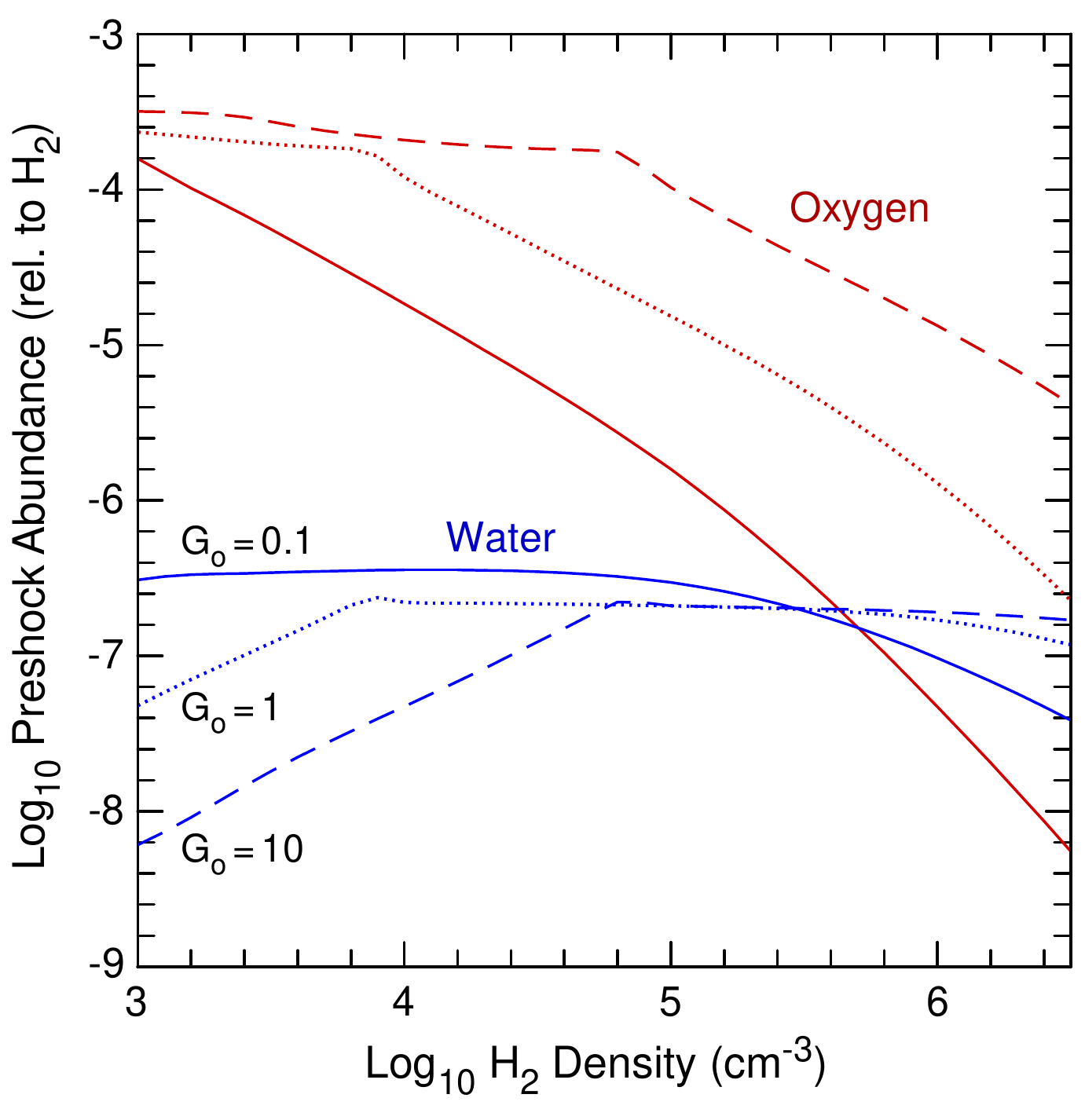}
\vspace{4mm}
\caption{Preshock gas-phase abundances of atomic oxygen (red curves) and \water\ (blue curves) as a function of 
preshock \mh\ density and FUV field strength, \go.  The abundances are determined from the PDR model 
used here (see text) assuming a visual extinction \av$=$1.0 between the FUV source and the preshock gas.
For a given density, lower values of \go\ result in greater amounts of O incorporated into ice mantles 
and thus not available in the gas phase. In the absence of sputtering, the preshock gas phase O abundance sets the 
maximum postshock \water\ abundance.}
\renewcommand{\baselinestretch}{0.92}
\vspace{3mm}
\label{preshockfig}
\end{figure}

To determine the conditions in the preshock molecular gas, we use the detailed 
photodissociation region (PDR) model of \citet{Hollenbach09}. This model computes the abundances of numerous 
atomic and molecular species, as well as those of charged and neutral dust grains and PAHs, as a function 
of extinction into a cloud, 
given the input parameters of gas density, $n$, and FUV radiation field strength, \go\
(scaling factor in multiples of the average Habing local interstellar radiation field; \citealt{Habing68}).
In order to model a shock propagating through such gas, the important outputs of the PDR model are: 
(i) the types and abundances of charge carriers, which determine how well the ionized species couple to the magnetic 
field; and (ii) the abundances of gas-phase species, especially oxygen- and carbon-bearing species that become 
incorporated into important gas coolants. It is the interplay between momentum transfer (moderated by the 
coupling length) and the efficiency of gas cooling (which limits the gas temperature) that determines the final 
shock structure. As shown by \citet{Hollenbach09}, significant fractions of oxygen and carbon nuclei can be locked 
up on the surfaces of dust grains in molecular gas exposed to moderate FUV fields at extinctions \av$\,\sim\,$1.
Based on the same PDR model described in detail in \citet{Hollenbach09}, Fig.~\ref{preshockfig} shows the
preshock gas-phase abundances of \water\ and O as functions of density and \go.

To first order, the length scale depends only on the ionization fraction and the strength of the magnetic field.
In gas fully-shielded from FUV radiation, the fractional ionization is set by the cosmic ray ionization rate and is low,
resulting in a relatively long ion-neutral coupling length scale and $C$-shocks that can be supported up to 
velocities of $\sim\,$40\kms.  In unshielded gas, the fractional ionization reaches 
its maximum value of $\sim\,$10$^{-4}$, set by the carbon abundance, and results in the shortest possible length 
scale. In gas at \av$\,=\,$1, the length scale is between these two extremes and depends sensitively on the gas 
density and grain properties.  An important point arising from these calculations is that, because the length scale 
is shorter than in fully-shielded gas, the maximum $C$-shock velocity is generally less than 40\kms.

\vspace{1mm}

\subsection{\em Postshock Chemistry}

\vspace{-1mm}

Within the shock, the abundances of gas-phase O-bearing species are set by competition between neutral-neutral 
reactions, which tend to drive O nuclei into \oxy, OH and \water, and photodissociation, which destroys these 
molecules. Studies of $C$-type shocks in fully-shielded gas show that O-nuclei are efficiently driven 
into \water\ once 
the gas temperature exceeds $\sim\,$400$\:$K, at shock velocities $\simgt\,$10\dash 15\kms; in the absence of 
photodissociation, the \water\ abundance remains high in the postshock gas.  The presence of FUV radiation 
has two effects.  First, it shortens the length scale and thus lowers the shock velocity at which the gas 
reaches 400$\:$K.
Second, it destroys molecules in the cooling postshock gas; to account for this, we have added photodissociation 
reactions to the shock chemical scheme, including dissociation of \water, OH and \oxy. Under these conditions, 
the gas has higher abundances of O and OH than without photodissociation.  The O and OH are driven into molecular oxygen by the reaction O$\:+\:$OH$\:\rightarrow\:$\oxy$\:+\:$H, which has a reaction rate that is within a 
factor of two of 3\ti 10$^{-11}$ cm$^3$ s$^{-1}$ over the range 10$\:$K to 500$\:$K \citep{McElroy13}.
This is in contrast with the neutral-neutral reactions that drive O into OH and \water, each of which has barriers 
of $>\,$1000$\:$K.  We find that the postshock \oxy\ abundance is directly correlated with the preshock 
gas-phase O-nuclei abundance; more O-nuclei in the preshock gas lead to more \water\ in the warm gas followed
by more \oxy\ in the cooler gas.  

In addition to increases in the preshock gas-phase O-nuclei abundance caused by FUV radiation, \water\ can
be removed from ice mantles by the shocks themselves.
Ice mantles on grains are sputtered by ion-neutral collisions in shocked regions where there is substantial relative
velocity between charged grains and neutral molecules. We follow the treatment of \citet{Flower94} for the removal
of oxygen and water-ice mantles, but with a threshold energy for ice removal consistent with that found by 
\citet{Neufeld14}, i.e.~complete grain mantle removal for shocks with $v_s\,>\:$25\kms. Note that these results are
similar to those found by both \citet{Draine83b} and
\citet{Jimenez-Serra08}, who find $\sim\,$10\% of mantles are removed in 20\kms\ shocks. 
At velocities less than $\sim\,$20\kms, sputtering of grain mantles is not expected to be efficient,
so the gas-phase abundances of O and C nuclei should be unchanged by the passage of such slower shocks.

Because the \oxy\ itself is subject to  photodissociation, 
the \oxy\ abundance, $N$(\oxy)/$N$(\mh)$\;=\:$\aoxy, does not reach 
the abundance of O-nuclei in the gas.  An example may be seen Fig.~\ref{gocomparisons} in which
models with \go$\,=\,$0.1, 1 and \go$\,=\,$10 are compared.  Note that the peak \oxy\ abundance is higher 
in the \go$\,=\,$10 case since the higher \go\ keeps more O in the gas phase.
Moreover, the higher postshock gas temperatures when
\go\ is high allow the neutral-neutral reactions to more than compensate for the photodestruction of \water\ while
$T\,\simgt\:$300$\:$K, keeping the \water\ abundance high; however, when $T$ drops below $\sim\,$300$\:$K, 
the \water\ photodestruction rate exceeds its formation rate and the \water\ abundance quickly drops, 
as shown in Fig.~\ref{gocomparisons}.

The conditions necessary to provide a significant \oxy\ column are: (i) a large fraction of O-nuclei in the preshock gas;
(ii) sufficient FUV radiation to dissociate the shock-produced \water, but not so much that no 
significant \oxy\ survives in the postshock gas; and (iii) a FUV flux that remains below the threshold at which
the length scale will no longer support $C$-shocks. 


\vspace{1mm}

\section{MODELING APPROACH}

\vspace{1mm}

\subsection{\em Shock Geometry}

\citet{Chen14} have identified Orion \mh\ Peak 1 as the most likely source of the reported
\oxy\ emission.  Peak 1 is the site where the high-velocity gas from the dynamical center of the outflow, about
30\asec\ to the southeast in the vicinity of the BN/KL region, 
impacts the surrounding cloud, creating strong and extended collisionally-excited
near-infrared \mh\ emission \citep{Cunningham06} as well as 22~GHz \water\ maser emission
\citep[cf.][]{Genzel81}.  Using
{\em Herschel}/HIFI, \citet{Neufeld13} recently reported the detection of \water\ maser emission arising from 
the 5$_{32}$\dash 4$_{41}$ rotational transition at 620.701~GHz toward \mh\ Peak 1 (see Fig.~\ref{maserfig}).  
As noted in Chen et al., the similarity between the LSR velocities of the maser emission of $\approx\,$11\kms\
and that of the \oxy\ suggests a connection.

Along with 22~GHz \water\ maser
observations obtained during the same epoch, Neufeld et$\,$al.~interpret the maser emission
as being consistent with a pumping model in which the population inversions arise from the
combined effects of collisional excitation and spontaneous radiative decay.  Moreover, because
collisional excitations that produce a population inversion in the 5$_{32}$\dash 4$_{41}$ transition
require \mh\ densities greater than 10$^8$~\cmc\ \citep[cf.][]{Neufeld91},
and because the upper 5$_{32}$ state lies 698$\:$K above the ground state, gas compressed and
heated behind a shock front provides the most plausible explanation for the source of this maser emission.

Finally, the direction of maximum gain for stimulated emission lies along the path with the minimum 
velocity gradient which, for a $C$-type shock, is perpendicular to the direction of propagation.  Thus, it's likely 
that the shocks giving rise to the 22~GHz and 621~GHz \water\ 
masers toward \mh\ Peak 1 are propagating perpendicular to our line of sight.  
While we are not suggesting that the \oxy\ emission necessarily arises from the \water\ maser spots,
the presence of the \water\ masers points to shocks moving in the plane of the sky, some of which
achieve high enough postshock densities and temperatures to produce maser action, but
most of which may not.  Such a scenario has 
significant advantages for explaining the \oxy\ emission observed toward \mh\ Peak~1: 
(1) shocks viewed in cross section (vs. face-on)
can potentially present the observer with a large column of gas with a relatively high postshock \oxy\
abundance; and, (2) the propagation velocity of the shocks in the plane of the sky can assume any value (up to the 
$C$-type shock breakdown velocity), while the line-of-sight velocity gradient can be low, and thus consistent
with the narrow ($\simlt\,$3\kms) \oxy\ line widths observed (see Table~1).  
This geometry is shown schematically in Fig.~\ref{shockgeo}.

\vspace{1mm}

\subsection{\em Shock Models and \oxy\ Emission}

A series of shock models were generated for preshock densities ranging between 10$^3\:$\cmc\ and 
10$^6\:$\cmc\ in steps of 0.1 dex and incident FUV fluxes, \go, of 0.1, 1, and 10.  
For each value of \go\ and preshock density, shock velocities ranging from
5\kms\ up to the $C$-shock breakdown velocity, in steps of 1\kms, were also generated.  In total,
839 shock models were computed for a \go$\,=\:$0.1, 618 models were computed for a \go$\,=\:$1, and
455 models were computed for a \go$\,=\:$10.  That fewer models were generated as
\go\ increases reflects the fact that as \go\ increases, $C$-type shocks break down at lower shock velocities.
Thus, a smaller range of shock velocities -- and models -- are consistent with $C$-type shocks
as \go\ rises.

For each shock model, the postshock \mh\ density, temperature, and \water, OH, and \oxy\ abundance
are computed as a function of distance behind the shock front.  The range of postshock 
distances considered extends beyond where: (1) the \water, OH, and \oxy\ abundances peak and 
subsequently drop by at least a factor of 10; and, (2) the postshock temperature
drops to 10$\:$K.  Meeting these criteria ensures that our modeling captures essentially all of
the \oxy\ emission since, beyond the postshock distances we consider, the reduced \oxy\ abundances combined 
with the low gas temperatures produce negligible contributions to the total \oxy\ integrated intensity.

Within the postshock region in which the density, temperature and \oxy\ abundance favor
non-negligible \oxy\ emission (as defined above), the gas is divided
into 1000 equally-spaced zones, each characterized by an \mh\ density, temperature, and \oxy\
abundance (relative to \mh) as computed by the $C$-shock code.
To obtain the \oxy\ line intensities from each zone, the equilibrium
level populations have been calculated using an escape probability method.  We use the
rate coefficients for collisions between \oxy\ and He computed by \citet{Lique10}, multiplied by 1.37 to
account for the different reduced mass when \mh\ is the collision partner.  Since these rates were
computed for gas temperatures $\leq\:$350$\:$K, extrapolation to higher temperatures was
necessary for application to postshock conditions.    To do this, a polynomial was fitted to the 
13 computed rates between 5$\:$K and 350$\:$K
for each transition and extended to 1400$\:$K.  As presented in \S4.3 (and figures therein), \oxy\
is formed in the postshock gas downstream of where \water\ and OH are formed.  At the postshock
distance where the \oxy\ abundance approaches its peak, gas temperatures have typically 
dropped below about 500$\:$K.  Thus, rates up to 1400$\:$K suffice.

Because \oxy\ has no dipole moment and can only emit quadrupole radiation, the line center optical
depth is small under almost all circumstances.  As a result, the emergent line intensity is proportional
to the \oxy\
column density, \noxy.  The range of \mh\ densities and \oxy\ abundances over which the postshock \oxy\ 
emission peaks is determined from the postshock densities, temperature, and abundances computed for 
each shock model, combined with the radiative transfer calculations; 
however, \noxy\ also depends on the line-of-sight depth of the shock, which is difficult to assess.  A study
of water maser emission from behind $C$-type shocks by \citet{Kaufman96b} suggests that the aspect ratio
of the zone of \water\ emission -- i.e., the ratio of the line-of-sight shock depth to the cross-sectional 
width of the high-\water\ abundance zone
on the plane of sky -- could exceed a factor of 100.  For our purposes, we assume the shock width (in the
plane of the sky) equals the distance between where the postshock gas temperature first rises above 10$\:$K
to where it first drops back to 10$\:$K.  We then multiply this distance by 100 to obtain the line-of-sight
depth of the \oxy\ emitting gas.  However, our analysis is not tied to the \mh\ and \oxy\ column densities
thus derived since the total integrated intensity measured for each \oxy\ line depends not only on \noxy,
but also on the number of shocked regions within the {\em Herschel}/HIFI beams, and hence the total 
spatial extent of the \oxy\ emission within these beams.   The spatial extent of the \oxy\ emission was 
only weakly constrained by 
\citet{Chen14}, other than inferring that the \oxy\ emission
most likely did not fill the {\em Herschel}/HIFI beams at 487, 774, and 1121~GHz.  Thus, we treat
the spatial extent of the \oxy\ emission as a free parameter.

We seek to match {\em both} the absolute \oxy\ line fluxes and the observed line ratios.  We
first derive the column densities and emission area needed to reproduce the measured \oxy\ 487~GHz 
line flux.  The ratio of the accompanying line fluxes at 774~GHz and 1121~GHz to that at 487~GHz
are then compared with those observed.
We assume that the \mh\ and \oxy\ column densities and emission area needed to reproduce the observed
487~GHz \oxy\ line flux are given by:

\vspace{-7mm}

\begin{eqnarray}
N({\rm H_2}) & = & 100 \times n({\rm H_2})\,d_{10}\,\gamma  \\*[3mm]
N({\rm O_2}) & = & N({\rm H_2})\,\chi({\rm O_2})~~,
\end{eqnarray}

\vspace{2mm}

where 100 is the assumed aspect ratio, $n$(\mh) is the \mh\ postshock density in the zones where 
487~GHz emission peaks, $d_{10}$ is the postshock distance between where the gas temperature 
first rises above 10$\:$K and where it first returns to 10$\:$K, and \aoxy\ is the \oxy\ abundance
relative to \mh.  Because the \oxy\ column density required to reproduce the observed
optically thin line flux is inversely proportional to the area of the emitting region, we introduce
the scaling factor $\gamma$, which is defined as

\vspace{-5mm}

\begin{equation}
\gamma = \frac{400~{\rm sq.~arcsec}}{A_{487}}~~,
\end{equation}

\vspace{3mm}

where 400 sq.~arcsec is chosen to be roughly consistent with the area of the \oxy\ emitting region assumed by
\citet{Chen14}, and $A_{487}$ is the actual area of the \oxy\ 487~GHz emission in sq.~arcsec were it 
known.  Since we compute the 487, 774, and 1121~GHz line emission as a function of postshock
depth, the area of the 774~GHz and 1121~GHz emitting regions, relative to $A_{487}$, is computed for 
each shock model and scale as

\vspace{-4.8mm}

\begin{eqnarray}
{\rm Relative~area~of~774~GHz~emitting~region} & = & A_{487}\;\left(\frac{w_{774}}
{w_{487}}\right)  \\*[3mm]
{\rm Relative~area~of~1121~GHz~emitting~region} & = & A_{487}\;\left(\frac{w_{1121}}
{w_{487}}\right)
\end{eqnarray}

\vspace{3.8mm}

where $w_{487}$, $w_{774}$, $w_{1121}$ are the widths of the 487~GHz, 774~GHz, and 1121~GHz
emitting regions behind the shock front, respectively.  The width of the emitting region in each line
is taken to be the range of postshock depths over which the line flux remains greater 
than 1 percent of its peak value.

Finally, we define a `Goodness of Line Ratio Fit' parameter to measure how closely each model comes
to reproducing the observed 774~GHz and 1121~GHz line fluxes once the 487~GHz line flux has been matched:

\vspace{1mm}

\begin{equation}
{\rm Goodness~of~Line~Ratio~Fit} = 
\left|\frac{R_{\rm 774m} - R_{\rm 774o}}{R_{\rm 774o}}\right| + 
\left|\frac{R_{\rm 1121m} - R_{\rm 1121o}}{R_{\rm 1121o}}\right|
\end{equation}

\vspace{5.9mm}

where $R_{774\rm m}$ is the model-predicted ratio of the 774$\:$GHz integrated intensity to the
487$\:$GHz integrated intensity, and $R_{774\rm o}$ is the observed ratio of the 774$\:$GHz integrated 
intensity to the 487$\:$GHz integrated intensity.  Likewise, $R_{1121\rm m}$ is the model-predicted 
ratio of the 1121$\:$GHz integrated intensity to the
487$\:$GHz integrated intensity, and $R_{1121\rm o}$ is the observed ratio of the 1121$\:$GHz integrated 
intensity to the 487$\:$GHz integrated intensity.  \citet{Chen14} report an observed 487\,:\,774\,:\,1121~GHz 
integrated line intensity ratio of 1\,:\,1.90\,:\,0.53 toward Orion \mh\ Peak 1.  As indicated by
Eqn.~(6), model-predicted line ratios that 
exactly match the observed ratios would result in a `Goodness of Line Ratio Fit' of zero.

\vspace{1mm}

\subsection{\em Shock Models Results}

The results for shock models of varying preshock density, shock velocity and FUV field, computed for 
\go$\,=\:$0.1, 1, and 10, are shown in Figs.~\ref{shockmodels0.1}, \ref{shockmodels1}, 
\ref{shockmodels10}, respectively.  Only those models that produce a `Goodness of Line Ratio Fit' less 
than 6 for \go$\,=\:$0.1 and 1, and less than 14 for \go$\,=\:$10, are shown for clarity.   The
best-fit model for each value of \go\ is summarized in Table~2; the \oxy\ line ratios produced
by these best-fit models are shown in Fig.~\ref{ratioplot}.
The profiles of postshock \oxy\ line integrated intensity, 
abundance, and temperature for the best-fit \go$\,=\:$0.1, 1, and 10 shocks models are
shown in Figs.~\ref{shockconditions01}, \ref{shockconditions1}, and \ref{shockconditions10}, respectively.
Finally, the line center optical depth, $\tau$, in the 487, 774, and 1121~GHz transitions was evaluated
for each of the best-fit results given in Table~2.  In no case did $\tau$ exceed 0.07 and, in most cases,
was significantly lower, thus justifying the optically thin assumptions made here.

\vspace{1mm}

\section{DISCUSSION}

\vspace{1mm}


Following more than six years of observations, and more than 7000 Galactic lines of sight surveyed,
the {\em SWAS} mission reported only one tentative detection of \oxy, toward the Rho Ophiuchi
molecular cloud \citep{Goldsmith02}.  Likewise, the {\em Odin} mission reported only upper limits
to the \oxy\ line strengths \citep{Pagani03, Sandqvist08}, with the exception of one possible detection,
also toward the Rho Ophiuchi cloud \citep{Larsson07}.  The upper limits to the \oxy\ abundance
set by {\em SWAS} and {\em Odin} are more than 100 times lower than that predicted by equilibrium
gas-phase chemical models.  This discrepancy was later understood to be primarily the result of the 
exclusion of dust grains from these models, which serve as important sites for both the freeze out 
of \water\ and the surface formation
of \water.   This water ice subsequently remains locked on grain surfaces until either photodesorbed by
FUV or X-ray photons, sublimated at grain temperatures above $\sim\,$100$\:$K,
or sputtered by shocks with velocities $\simgt\,$25~\kms\ 
\citep{Draine83b, Neufeld14}.  By sequestering large amounts of elemental oxygen in water ice,
\citet{Bergin00} and \citet{Hollenbach09} showed that the gas-phase production of \oxy\ is effectively
suppressed.  

The {\em Herschel} Oxygen Project, guided by these updated models, used {\em Herschel}'s 
greater sensitivity to continue the search for \oxy.  To date, however, Rho Ophiuchi \citep{Liseau12}
and Orion \mh\ Peak 1 remain the only sources with statistically significant \oxy\ detections.  
It might be expected that processes common to both sources, yet rare overall, could account for the
\oxy\ emission.
Unfortunately, the conditions inferred by Liseau et$\:$al.~to explain the
\oxy\ emission toward Rho Ophiuchi are insufficient to account for the emission toward Orion \mh\ Peak~1.
In particular, toward Rho Ophiuchi, the \oxy\ emission is attributed
to a combination of two emitting regions, one with \noxy$\:>\:$6\ti 10$^{15}$~\cmc\ and $T <\:$30$\:$K,
and the other with \noxy$\;=\;$5.5\ti 10$^{15}$~\cmc\ and $T >\:$50$\;$K.  The inferred 
beam-averaged \oxy\ abundance is $\sim\,$5\ti 10$^{-8}$ in the warmer component, and somewhat 
higher in the colder component.  The successful detection of \oxy\ toward this source, 
among the many sources toward which no \oxy\ emission was detected, 
was attributed to time-dependent quiescent cloud chemistry -- 
i.e.,  Rho Ophiuchi was surmised to have been observed during a relatively short period when the 
evolving \oxy\ abundance was near its peak.

Such a scenario is unlikely to apply to Orion \mh\ Peak 1.  First, the constraints on the
spatial extent of the \oxy\ emission toward Peak 1 provided in \citet{Chen14} -- i.e., an \oxy\ emitting
region less than 25\asec\ in diameter -- require that the \oxy\ column density be between
about 3\ti 10$^{17}$~\cms\ and 3\ti 10$^{18}$~\cms\ to produce the absolute \oxy\ line
intensities measured.  \oxy\ column densities this high are hard to produce in quiescent gas \citep[e.g.,][]
{Hollenbach09}.
Second, Peak 1 is the site of intense shock, rather than
quiescent cloud, emission.  Third, {\em if} a component of quiescent gas close to \mh\ Peak 1 were
responsible for the \oxy\ emission, and the \oxy\ abundance were close to its quiescent
cloud peak of $\sim\,$5\ti 10$^{-8}$, as
in Rho Ophiuchi, the required \mh\ column density, $N$(\mh)$\:=\:$\noxy\,/\,\aoxy,
 would be on the order of 10$^{25}$~\cms, which is much higher than observed.  Thus, 
process(es) different from those invoked to explain the \oxy\ emission toward Rho Ophiuchi are needed
to explain the detections toward Orion \mh\ Peak 1.

In addition to the presence of shock activity, Orion is the site of O and B stars that
produce strong FUV radiation.  The FUV field near the Trapezium has been estimated to be
\go$\,\simgt\:$10$^4$ based upon the total radiation from the Trapezium stars -- and the O star 
$\theta^1$ Ori C in particular.  The intensity of this field is corroborated by
the strength of the far-infrared [C\,II] and [O\,I] fine-structure
lines mapped toward the Orion molecular ridge, the strength of several near-infrared lines whose 
intensities have been ascribed to recombinations to highly excited states of
CI, and the strength of near-infrared N\,I lines excited by the fluorescence of UV lines 
\citep{Herrmann97, Marconi98, Walmsley00}.  The amount by which the FUV field is attenuated
between the Trapezium and \mh\ Peak 1 due to intervening material is uncertain,
as is the FUV radiation from other B stars in the Orion cloud.  However, as
discussed in \S3, even modest amounts of FUV radiation (i.e., \go$\,\leq\:$10) can affect both
the structure of and chemistry behind $C$-type shocks and motivate the study here.

FUV-illuminated $C$-type shocks provide a natural explanation for the Orion \mh\ Peak~1 \oxy\ emission 
for three reasons.  
First, as discussed in \S3.2 and \S3.3, the FUV field can increase the atomic oxygen abundance in the
preshock gas, which will increase the peak \water\ abundance in the postshock gas.
Second, the postshock OH abundance (and, ultimately, the \oxy\ abundance) is increased via
the photodissociation of \water\ as well as OH-producing chemical reactions, some with high activation barriers, 
enabled by the elevated postshock gas temperatures.  By raising the \oxy\ abundance above that attainable in
cold quiescent gas, the implied \mh\ column density can be brought closer to observed values.
Finally, the similarity between the LSR velocities measured for the \oxy\ lines of $\approx\,$11\kms\ 
(see Chen et al.), and those measured for the 22 GHz and 621~GHz \water\ masers of 10\dash 13\kms,
suggests a possible physical connection.
As discussed in \S 4.1, the maser emission indicates that some shocks are propagating in the 
plane of the sky, which can potentially provide a higher line-of-sight
column density of \oxy\ than face-on shocks while allowing for the narrow observed \oxy\ line widths.

For these reasons, as well as the conspicuous presence of shock activity associated with \mh\ Peak 1,
both \citet{Chen14} and we focus on shocks as the most likely source of the detected \oxy\ emission.
However, three significant differences distinguish the approach previously taken by Chen et\,al.~and that
taken here.  First, Chen et\,al.~assume that the preshock gas and the shock are fully shielded from FUV
radiation, with only the postshock gas subject to a FUV field.  Here, we assume that the preshock, shock,
and postshock gas are illuminated by a common FUV field.  Given that the width of the shock front is
approximately 10$^{16}$~cm or less (see Figs.~\ref{shockconditions01}, \ref{shockconditions1}, and
\ref{shockconditions10}), this approach seems well justified.  In particular, the analysis presented here avoids
the inconsistency in which the postshock physical conditions are determined assuming the absence of 
FUV radiation while the postshock chemistry, which is governed by these physical conditions (e.g., density and 
temperature), requires the presence of FUV radiation.

\begin{figure}[h]
\centering
\vspace{2mm}
$\!\!\!$\includegraphics[scale=0.77]{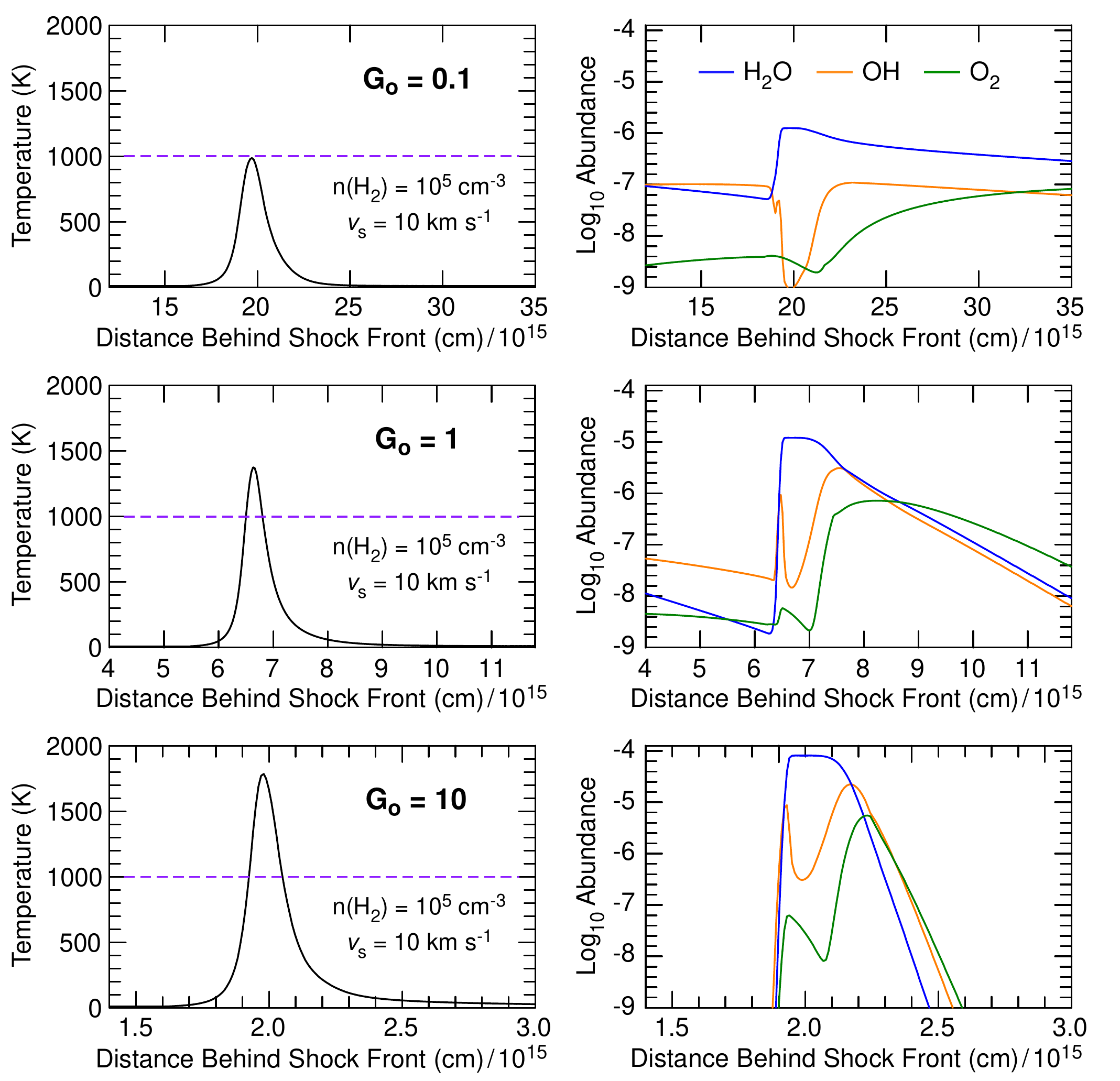}
\vspace{3mm}
\caption{Profiles of postshock gas temperature ({\em left panels}) and resulting \water, OH, and \oxy\ abundance
({\em right panels}) for \go$\,=\:$0.1 ({\em top row}), \go$\,=\:$1 ({\em middle row}), and \go$\,=\:$10 
({\em bottom row}).  All models have been run for  a preshock \mh\ density of 10$^5$~\cmc\ and a shock 
velocity of 10\kms.  Because this shock velocity is low, the preshock gas-phase O abundance is not increased
by sputtering.  The fiducial temperature of 1000$\:$K is denoted by a horizontal dashed line
in the left panels to highlight the increase in peak postshock temperature with increasing \go.  Also note that
distance behind the shock front at which the gas temperature and \water, OH, and \oxy\ abundances peak, 
and the length scale over which
the elevated temperatures and abundances persist, decrease with increasing \go.}
\renewcommand{\baselinestretch}{0.92}
\vspace{3mm}
\label{gocomparisons}
\end{figure}

\begin{figure}[h]
\centering
\vspace{-3mm}
$\!\!$\includegraphics[angle=90,scale=0.715]{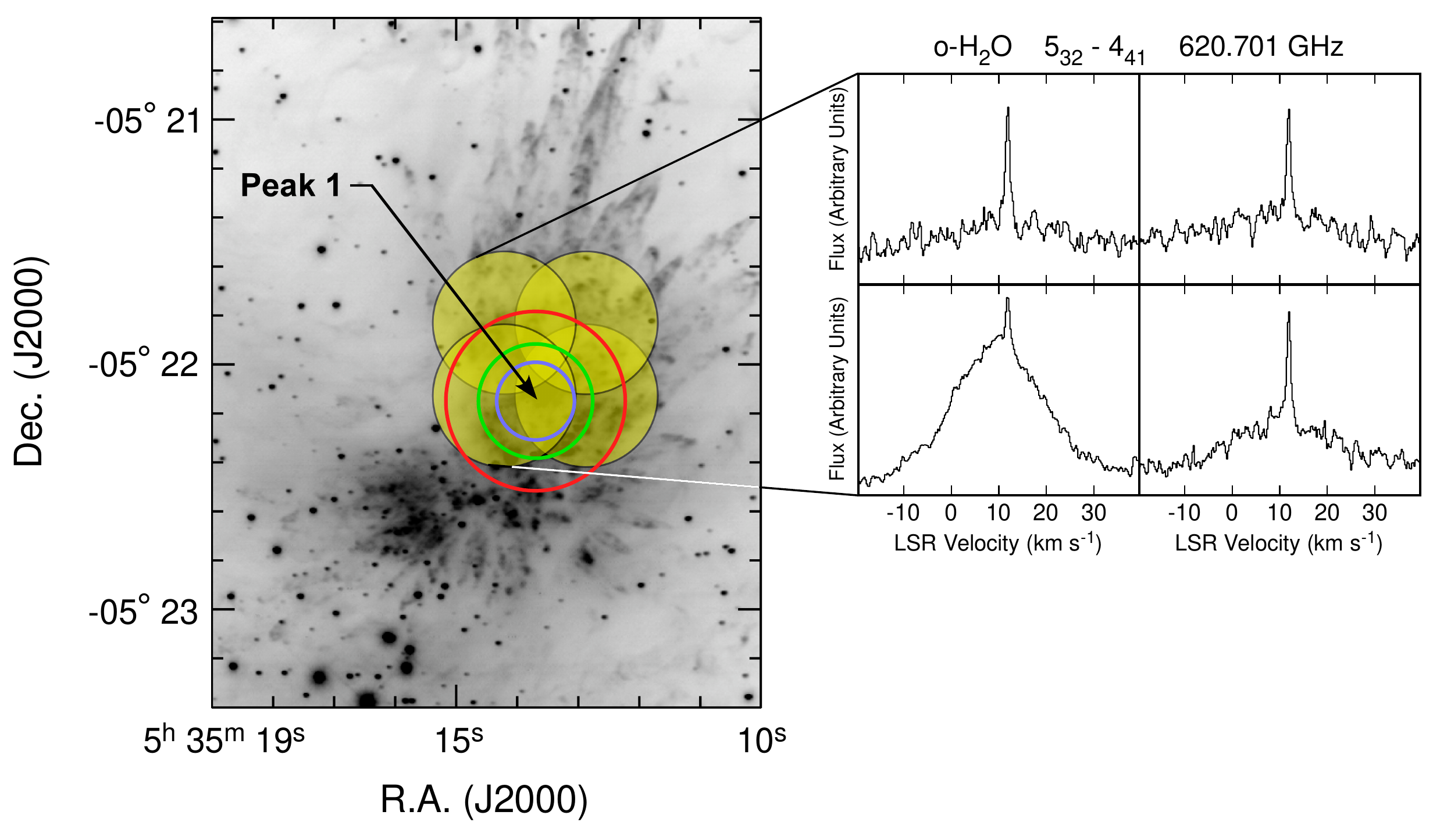}
\vspace{1.5mm}
\renewcommand{\baselinestretch}{1.04}
\caption{Position of \oxy\ 487 GHz ({\em red circle}), 774 GHz ({\em green circle}), and 1121 GHz ({\em blue circle}) 
beams centered on \mh\ Peak 1 ($\alpha_{2000} =\:$5$^{\rm h}$ 35$^{\rm m}$ 13$^{\rm s}\!\!.$7,
$\delta_{2000} =\:-$5$^{\rm o}$ 22\amin\ 09\asec; after Chen \etal\ 2014) superposed on the \mh\ image obtained
by \citet{Cunningham06}.  The {\em Herschel}/HIFI instrument
was also used to map a $\sim\,$2\amin\ \ti\ 2\amin\ region centered on BN/KL; only those positions toward
Peak~1 exhibit evidence for \water\ maser emission \citep{Neufeld13}.  The 621 GHz HIFI beams
and their sky positions are shown as filled yellow circles.  The \water\ spectra, with the prominent
maser feature at $\approx\,$11\kms, is shown on the right.
}
\renewcommand{\baselinestretch}{0.92}
\vspace{3mm}
\label{maserfig}
\end{figure}

\begin{figure}[h]
\centering
\vspace{13mm}
$\!\!$\includegraphics[scale=0.77]{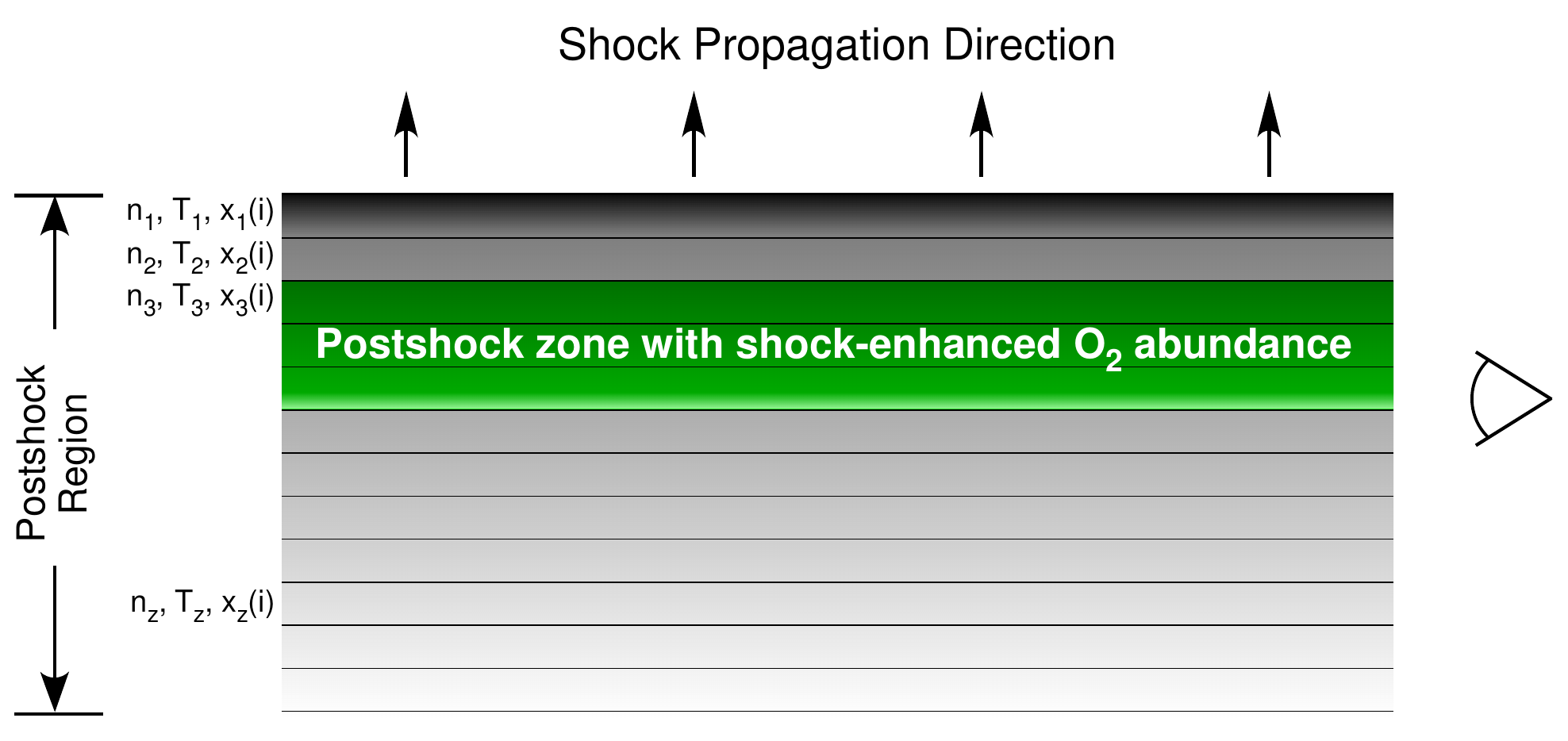}
\vspace{25mm}
\renewcommand{\baselinestretch}{1.04}
\caption{Proposed scenario in which the \oxy\ emission arises within gas in which the \oxy\ abundance is enriched 
by the passage of a non-dissociative $C$-type shock.  If these shocks are propagating perpendicular to our line of sight,
as suggested by the presence of \water\ masers toward Orion \mh\ Peak 1, the column density of warm, \oxy-enriched
gas can be significantly greater than for face-on shocks.  Such a scenario permits a line-of-sight
LSR velocity shift of $\sim\,$12~\kms, while the perpendicular shock velocity can be a free parameter.  Finally,
because the line of sight samples gas with a low velocity gradient, consistent with maser 
production, the measured \oxy\ line widths of $\leq\:$3\kms\ can be understood.  The models described here
divide the postshock region into 1,000 zones, each of which is characterized by a distinct density, temperature, and set of
chemical abundances.  The sum of the computed \oxy\ emission from all of these zones is then compared with the
observed integrated intensity in each of the \oxy\ lines detected.}
\renewcommand{\baselinestretch}{0.92}
\vspace{3mm}
\label{shockgeo}
\end{figure}

\begin{figure}[h]
\centering
\vspace{-3mm}
$\!\!$\includegraphics[scale=0.82]{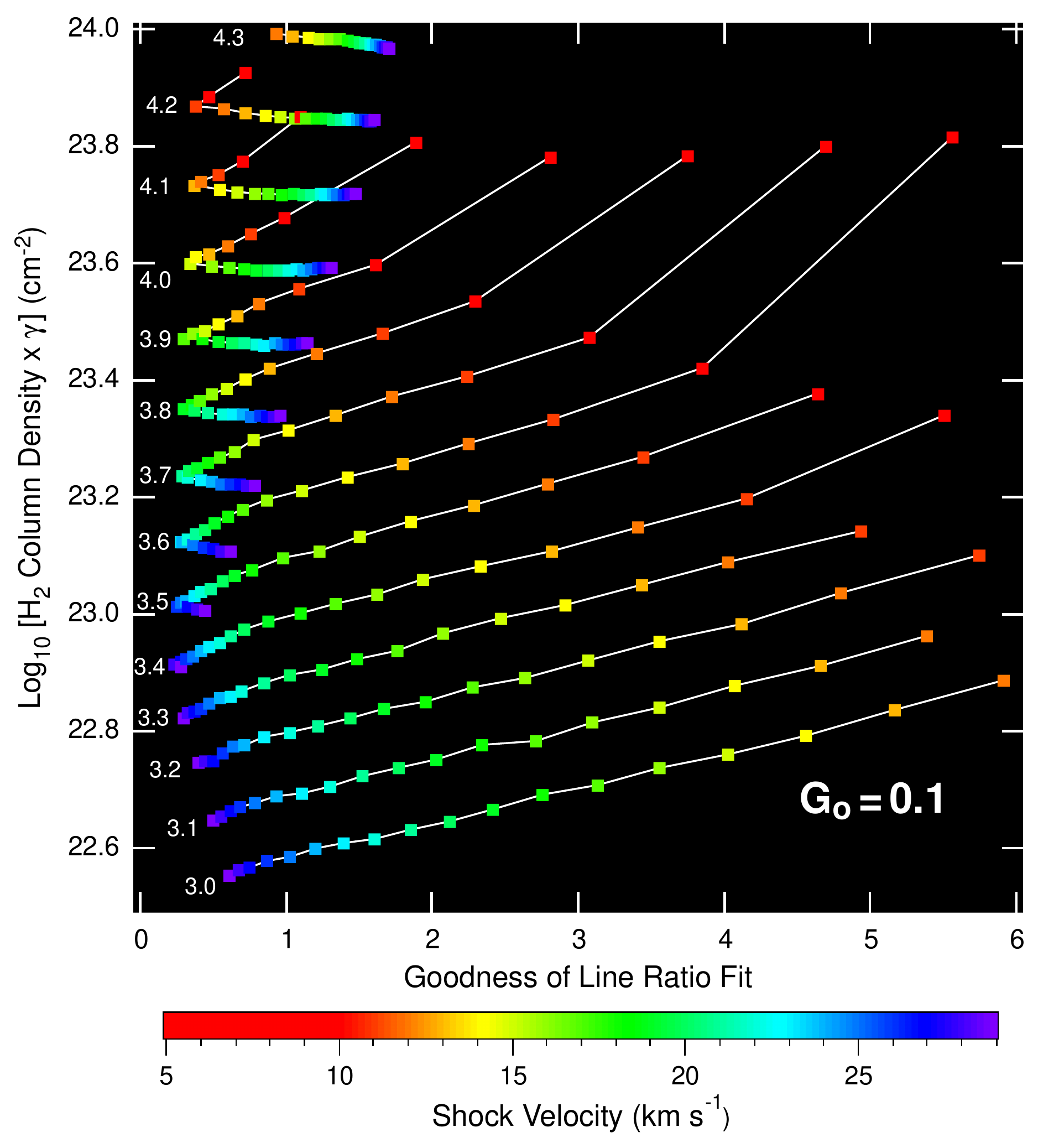}
\vspace{4mm}
\caption{Plot of the agreement between the predicted and observed 487 GHz, 774 GHz, and 1121 GHz \oxy\ line ratios 
produced by various shock models versus \mh\ column density and shock velocity
for \go$\:=\:$0.1.  As discussed in the text, the lower the value of the Goodness of Fit, the better the shock conditions
reproduce the observed \oxy\ line ratios, with a value of zero representing a perfect fit.  Log$_{10}$ of the preshock \mh\
density is shown for each family of models.  The required \mh\ (and \oxy) column density scales inversely with
the area of the \oxy\ emitting region.  The above plot assumes a 487 GHz \oxy\ emitting area of 400 sq.~arcseconds
(see text).}
\renewcommand{\baselinestretch}{0.92}
\vspace{3mm}
\label{shockmodels0.1}
\end{figure}

\begin{figure}[h]
\centering
\vspace{-3mm}
$\!\!$\includegraphics[scale=0.82]{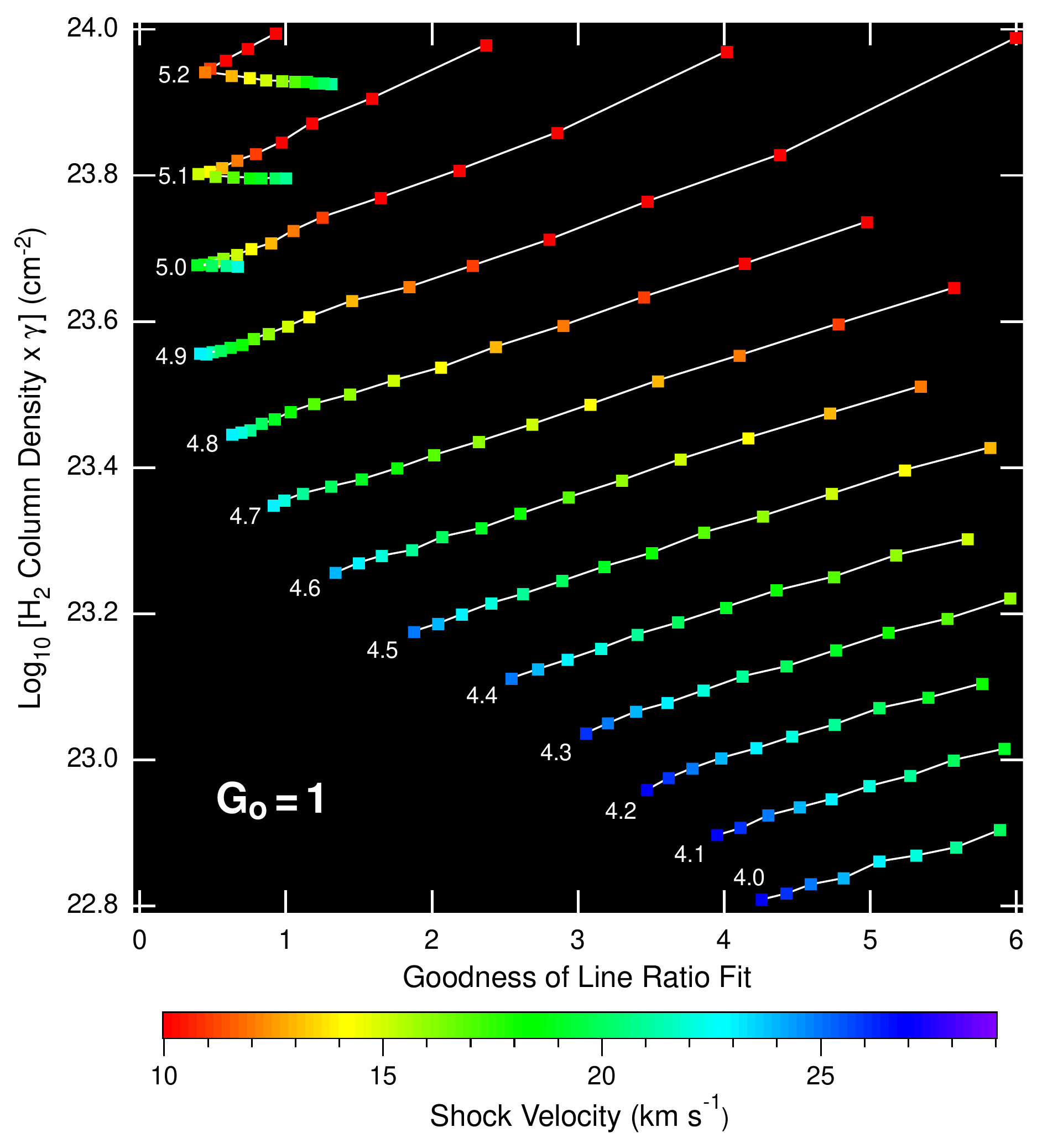}
\vspace{4mm}
\caption{Plot of the agreement between the predicted and observed 487 GHz, 774 GHz, and 1121 GHz \oxy\ line ratios 
produced by various shock models versus \mh\ column density and shock velocity
for \go$\:=\:$1.  As discussed in the text, the lower the value of the Goodness of Fit, the better the shock conditions
reproduce the observed \oxy\ line ratios, with a value of zero representing a perfect fit.  Log$_{10}$ of the preshock \mh\
density is shown for each family of models.  The required \mh\ (and \oxy) column density scales inversely with
the area of the \oxy\ emitting region.  The above plot assumes a 487 GHz \oxy\ emitting area of 400 sq.~arcseconds
(see text).}
\renewcommand{\baselinestretch}{0.92}
\vspace{3mm}
\label{shockmodels1}
\end{figure}

\begin{figure}[h]
\centering
\vspace{-3mm}
$\!\!$\includegraphics[scale=0.82]{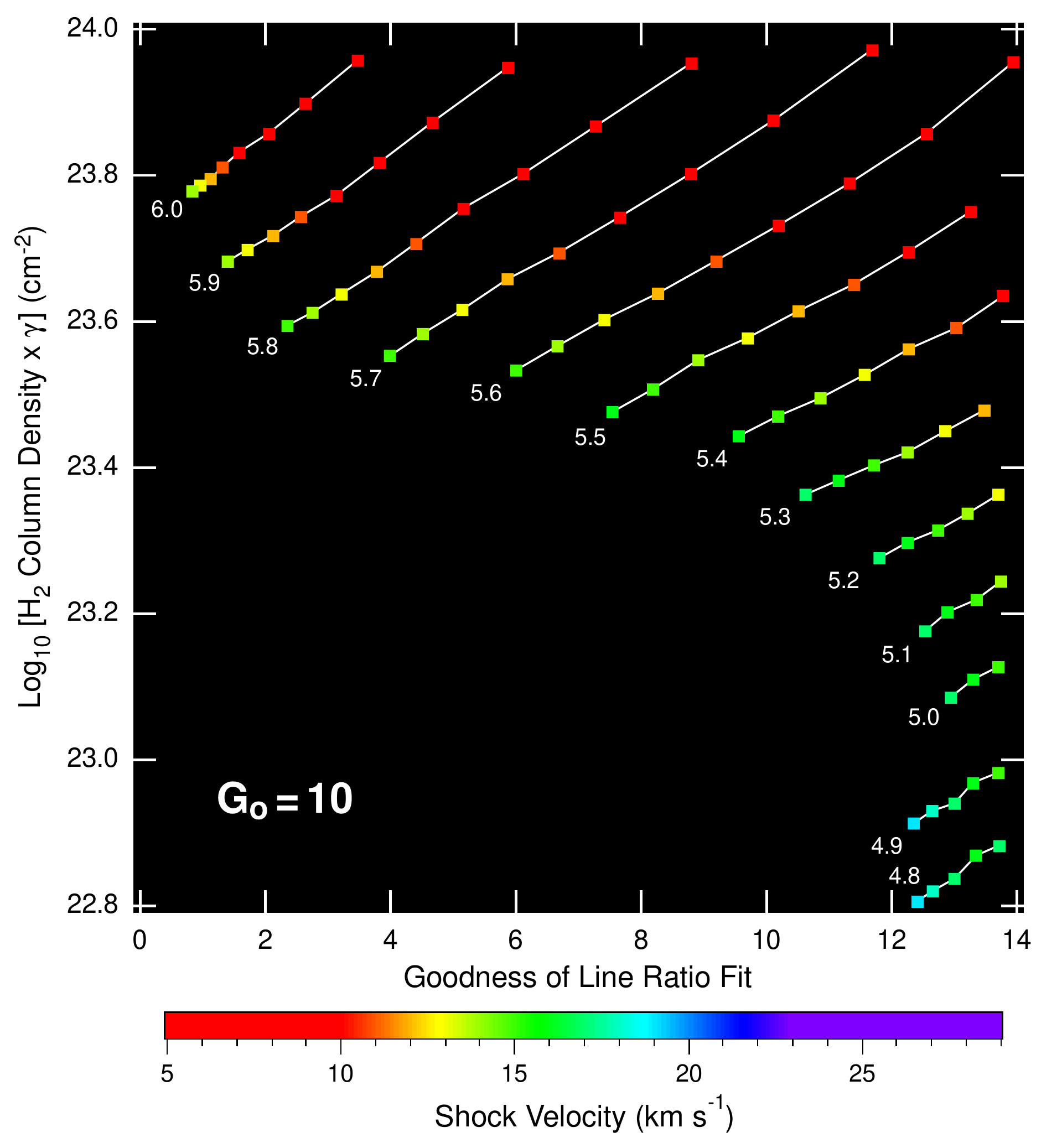}
\vspace{4mm}
\caption{Plot of the agreement between the predicted and observed 487 GHz, 774 GHz, and 1121 GHz \oxy\ line ratios 
produced by various shock models versus \mh\ column density and shock velocity
for \go$\:=\:$10.  As discussed in the text, the lower the value of the Goodness of Fit, the better the shock conditions
reproduce the observed \oxy\ line ratios, with a value of zero representing a perfect fit.  Log$_{10}$ of the preshock \mh\
density is shown for each family of models.  The required \mh\ (and \oxy) column density scales inversely with
the area of the \oxy\ emitting region.  The above plot assumes a 487 GHz \oxy\ emitting area of 400 sq.~arcseconds
(see text).}
\renewcommand{\baselinestretch}{0.92}
\vspace{3mm}
\label{shockmodels10}
\end{figure}

\begin{figure}[h]
\centering
\vspace{13mm}
$\:$\includegraphics[scale=0.79]{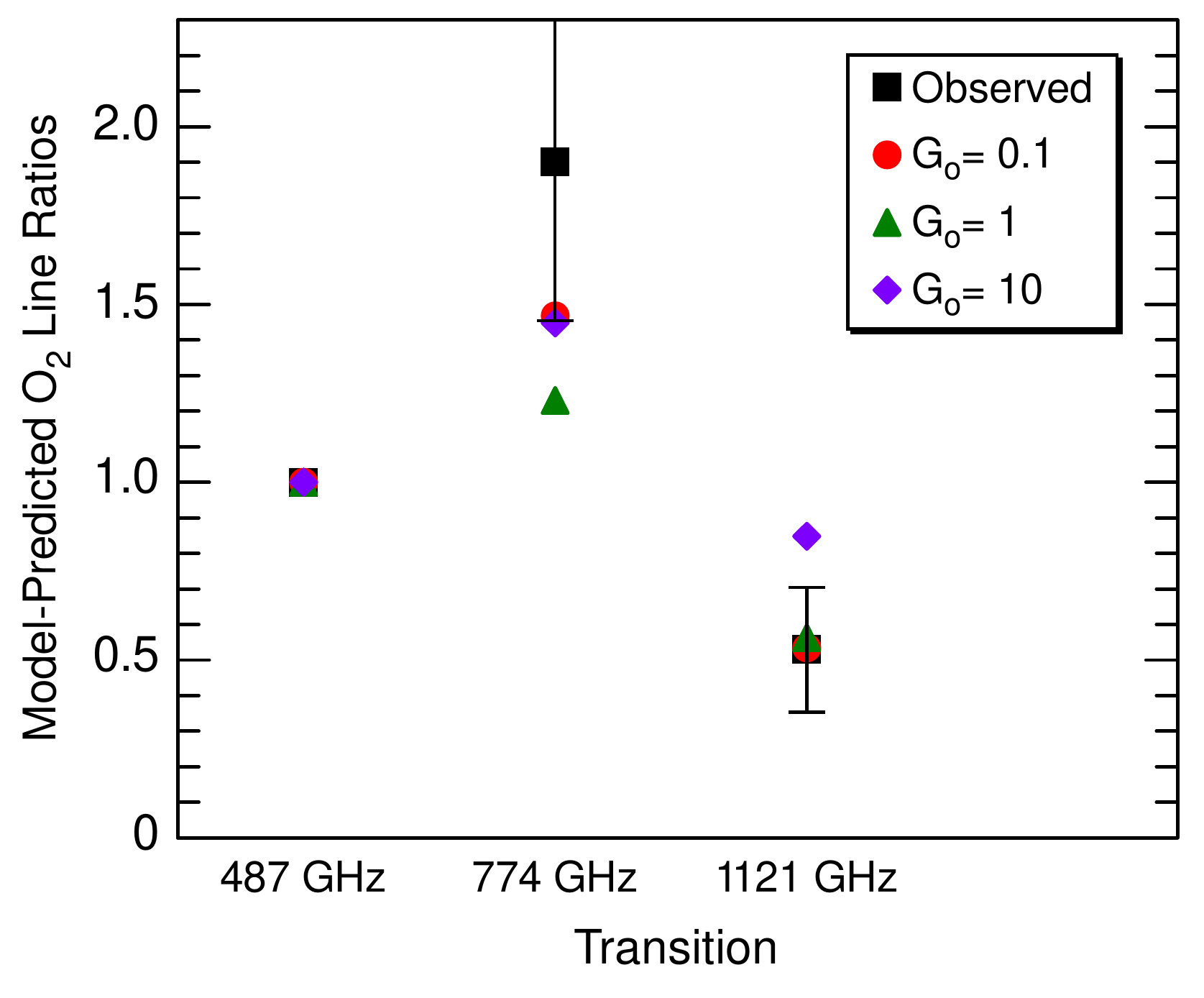}
\vspace{4mm}
\caption{Ratio of the observed \oxy\ integrated line intensities to that of the 487~GHz line ({\em black square})
given by \citet{Chen14}, and those predicted by the best-fit shock model with \go$\,=\:$0.1 ({\em red circle}),
\go$\,=\:$1 ({\em green triangle}), and \go$\,=\:$10 ({\em purple diamond}) and detailed in Table~2.
As discussed in the text, the shock models have been computed to reproduce the \oxy\ 487~GHz 
integrated intensity.  The 1$\sigma$ uncertainties in ratios of the 774~GHz and 1121~GHz line intensities
to the 487~GHz line intensity are shown.}
\renewcommand{\baselinestretch}{0.92}
\vspace{3mm}
\label{ratioplot}
\end{figure}

\begin{figure}[h]
\centering
\vspace{-3mm}
$\!\!$\includegraphics[scale=0.78]{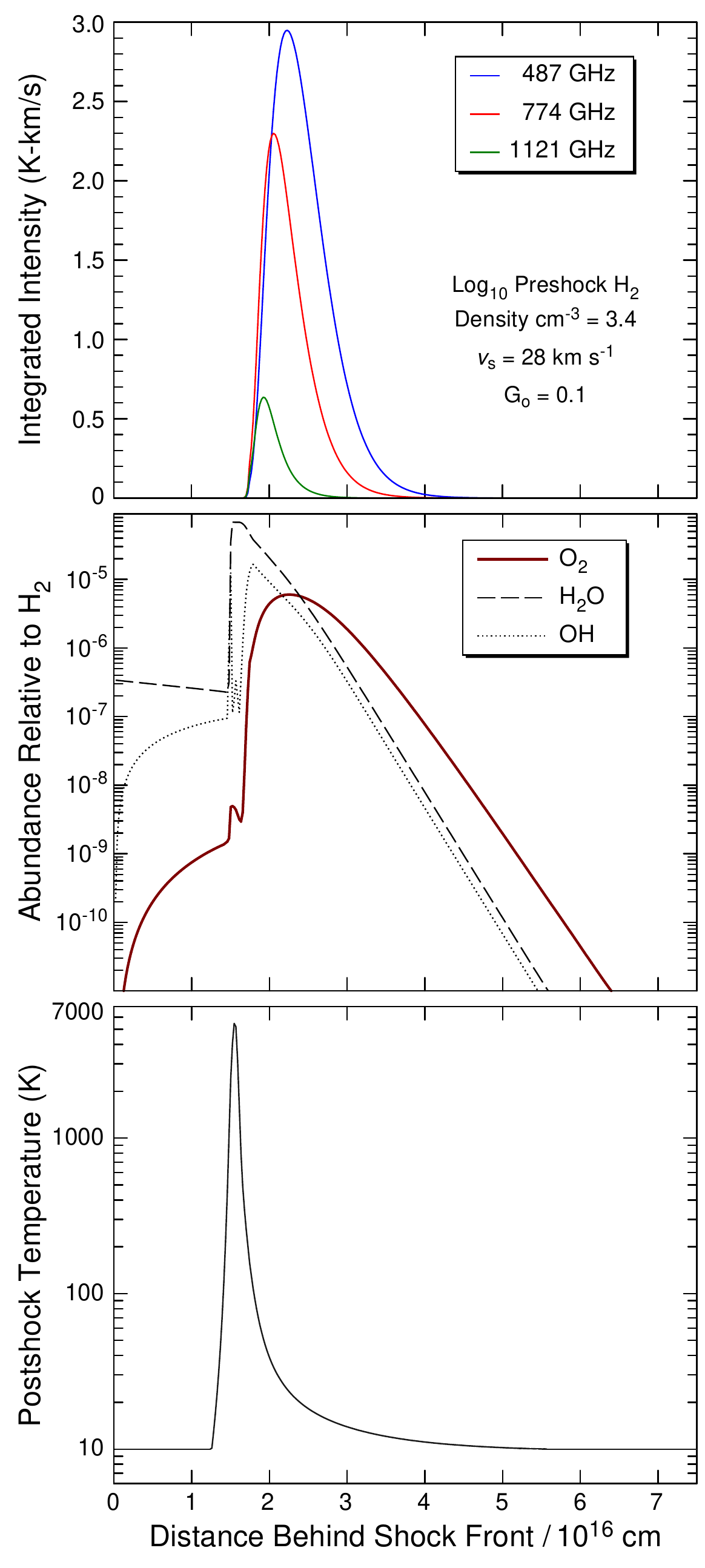}
\vspace{3mm}
\caption{Profiles of integrated intensity ({\em top}), abundance ({\em middle}), and gas temperature ({\em bottom})
as a function of distance behind the shock front for the best-fit shock parameters for \go$\:=\:$0.1.}
\renewcommand{\baselinestretch}{0.92}
\vspace{3mm}
\label{shockconditions01}
\end{figure}

\begin{figure}[h]
\centering
\vspace{-3mm}
$\!\!$\includegraphics[scale=0.78]{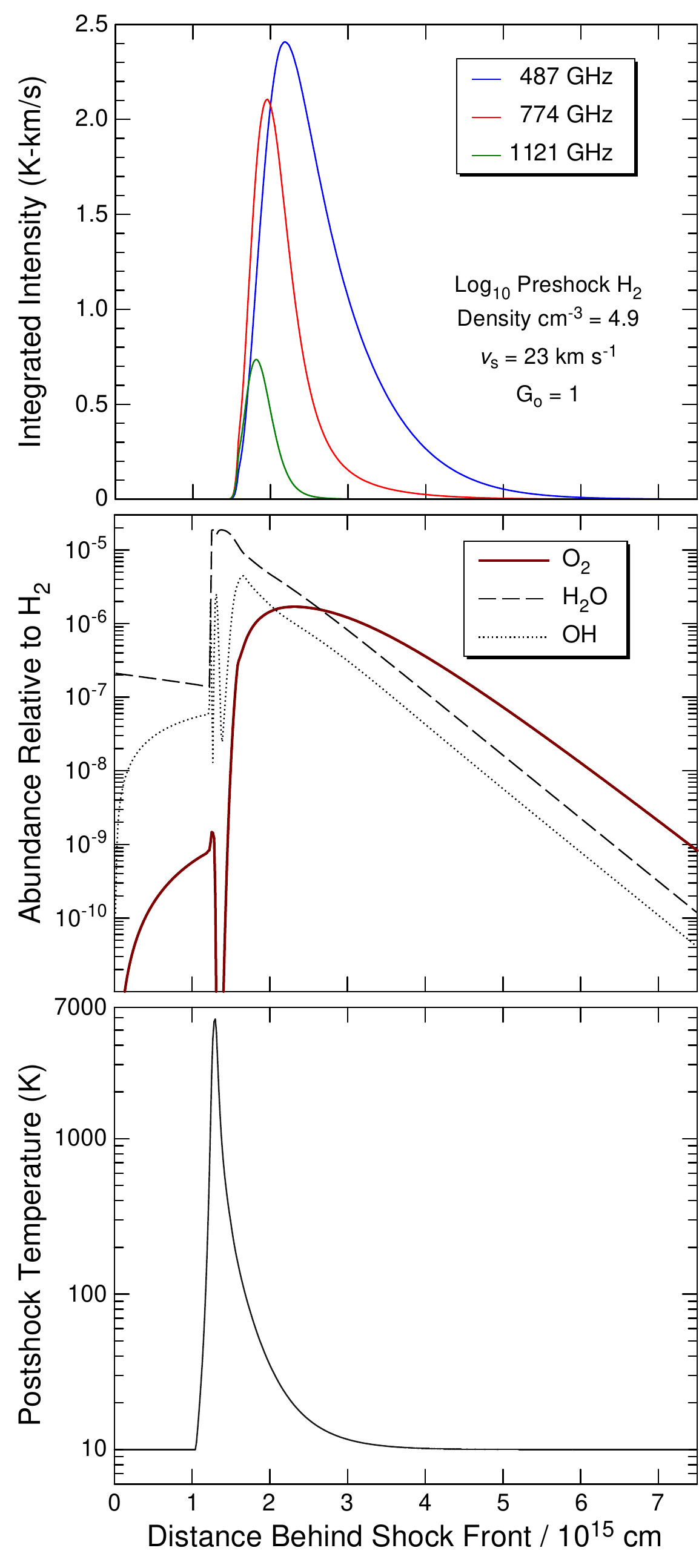}
\vspace{3mm}
\caption{Profiles of integrated intensity ({\em top}), abundance ({\em middle}), and gas temperature ({\em bottom})
as a function of distance behind the shock front for the best-fit shock parameters for \go$\:=\:$1.}
\renewcommand{\baselinestretch}{0.92}
\vspace{3mm}
\label{shockconditions1}
\end{figure}

\begin{figure}[h]
\centering
\vspace{-3mm}
$\:$\includegraphics[scale=0.78]{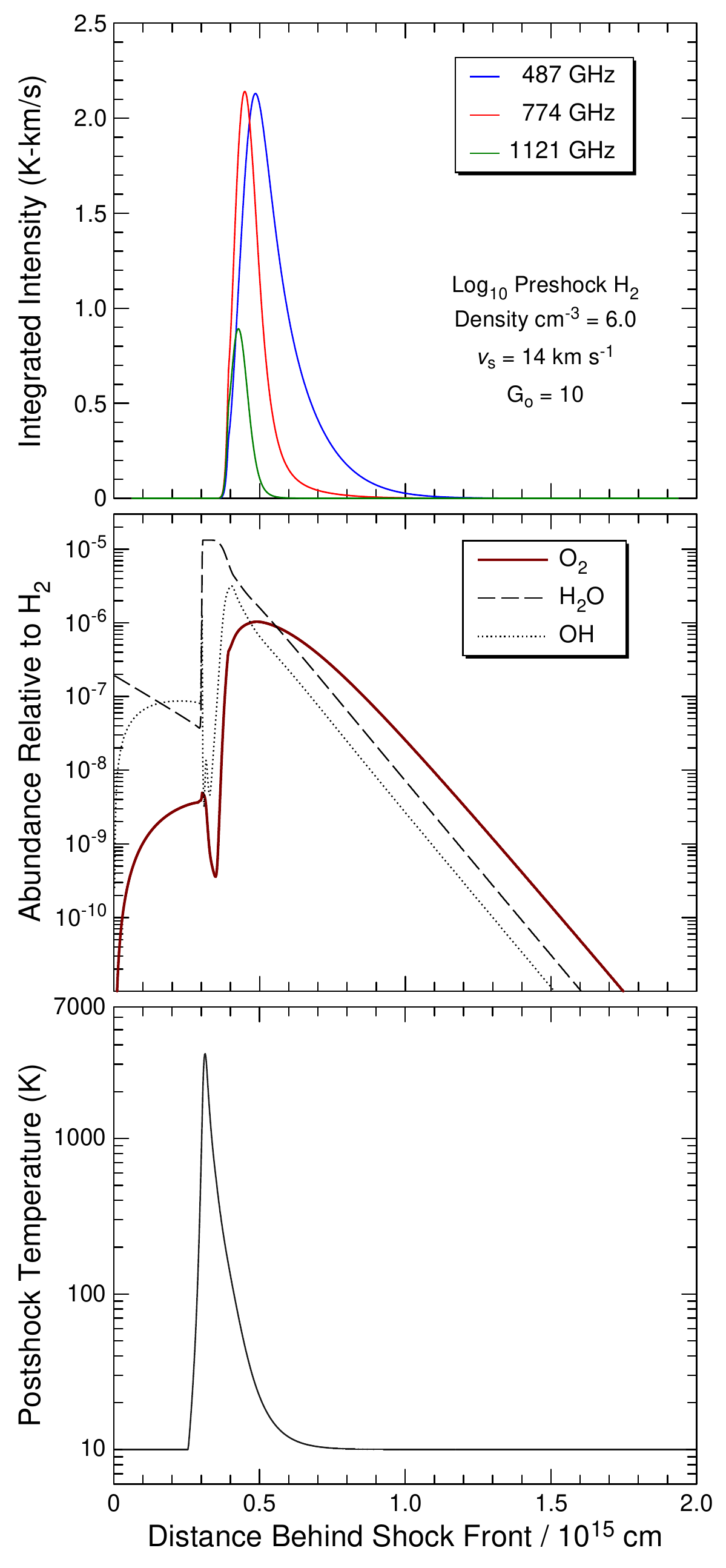}
\vspace{3mm}
\caption{Profiles of integrated intensity ({\em top}), abundance ({\em middle}), and gas temperature ({\em bottom})
as a function of distance behind the shock front for the best-fit shock parameters for \go$\:=\:$10.}
\renewcommand{\baselinestretch}{0.92}
\vspace{3mm}
\label{shockconditions10}
\end{figure}

Second, the above is important since the effects of FUV radiation on the shock structure can result in 
substantial changes to the shock width and postshock temperature, with the latter affecting the postshock chemistry.
As shown in Fig.~\ref{gocomparisons}, for a given preshock density and shock velocity, increasing the FUV field intensity
both increases the peak postshock temperature and reduces the shock width.   As shown in \citet{Bergin98}, the
formation rate of \water\ via the neutral-neutral reactions 
O$+$\mh$\,\rightarrow\,$OH$+$H; OH$+$\mh$\,\rightarrow\,$\water$+$H becomes important when the
postshock temperature exceeds $\sim\,$300$\:$K and increases rapidly with temperature.
In fact, as the peak postshock temperature rises with increasing \go, the \water\ formation
rate begins to exceed the \water\ photodestruction rate resulting in a net increase in the postshock
\water\ abundance with \go\ (e.g., compare Fig.~\ref{gocomparisons} middle and bottom panels).  This trend 
continues up to the full conversion into \water\ of all gas-phase O nuclei not locked in CO.  Only when the 
postshock cooling causes the temperature to drop below about 300$\:$K does the \water\ formation rate decrease 
significantly.  When this occurs, photodestruction dominates formation and the
\water\ abundance begins to decline, as shown in Fig.~\ref{gocomparisons}.   Ignoring the effects of FUV radiation
leads to an underestimate of the postshock temperature and the postshock depth to which \water\ exists in high abundance
as well as an overestimate of the shock width, all of which affect the resulting \oxy\ emission.

Third, since \oxy\ is formed from the photo-destroyed products of \water, i.e., O and OH, \water\ must be
relatively abundant (i.e., $\chi$(\water)$\,>\:$10$^{-5}$) in the postshock gas to produce the observed \oxy\ emission.
There are two ways to achieve this: (1) by placing \water\ into the gas phase directly through the sputtering of \water\
from the ice-covered mantles of dust grains behind the
shock; and, (2) by chemically forming \water\ behind the shock via the neutral-neutral reactions above.
However, for process (2) to form a high \water\ abundance, the preshock gas must have a high abundance 
of atomic oxygen.  Within gas in which the \water\ depletion onto grains is significant, Chen et\,al.~require
that the sputtering efficiency be $\simgt\:$10\% behind a 12\kms\ shock.  This efficiency exceeds all theoretical
predictions of which we are aware, including \citet{Draine83b}, \citet[][see their Figure 7]{Jimenez-Serra08}, 
and \citet{Flower10}.
Recently, \citet{Neufeld14} have modeled the water abundance behind interstellar shocks based on {\em Herschel}
measurements of far-infrared (IR) and submillimeter measurements of CO and \water\ in combination with
{\em Spitzer} measurements of mid-IR \mh\ rotational emission.  Their best-fit results are in good agreement
with the prediction that only when shocks reach a velocity of $\simgt\:$25\kms\ will \water\ be completely
removed from grain mantles.  Thus, it is not clear that sufficient \water\ can
be removed from grain mantles to support the model offered in Chen et\,al.~if depletion of \water\ onto 
dust grains is significant in the preshock gas
and negligible sputtering efficiency is assumed behind a 12\kms\ shock.

Alternately, Chen et\,al.~suggest that the preshock gas is sufficiently young that depletion is minimal,
in which case the atomic oxygen abundance is high.  However, since this model assumes the preshock gas is
collapsing from an initial density of 100$\;$\cmc, the time between when the density
evolves to its required preshock value (i.e., 4.2 \ti 10$^4\:$\cmc) and when depletion becomes important 
is $\sim\,$1.7 \ti 10$^6$ years.
Nevertheless, invoking a short interval during which the preshock is sufficiently dense, but undepleted,
could result in the high preshock O abundance needed to yield a postshock \water\ abundance sufficient to produce 
detectable \oxy\ emission.  A short-lived period in which both the preshock gas density and O abundance are high could
also account for the rarity of \oxy\ detections, as discussed by Chen et\,al.  However, this model still presumes 
the preshock gas is fully shielded while the postshock gas is not, an assumption not justified in the
Chen et\,al.~model.

Introducing a FUV field to the preshock gas solves these problems.  FUV photons can photodesorb \water\ from
grain mantles when shock velocities are too low to produce significant sputtering (i.e., $v_s\,\simlt\:$20\kms).  
(When shock velocities are $\simgt\:$25\kms, sputtering will remove any \water\ not photodesorbed
in the preshock gas.)  
Even if the FUV field results in the photodestruction of gas-phase \water\ in the 
preshock gas, the O and OH produced will be rapidly converted to \water\ in the warm postshock gas.
Thus, a FUV field can eliminate the need for high assumed sputtering efficiencies at low shock velocities while relaxing the
need to invoke a particular epoch in the time-dependent  evolution of the preshock gas.  As for time-dependent
preshock evolution
being the cause for the rarity of \oxy\ detections, a different explanation is suggested below.

Table~2 lists shock models considered here that both reproduce the absolute \oxy\ line intensities and line ratios 
for incident FUV fields having a \go$\:=\:$0.1, 1, and 10.
Several trends are evident.  First, as discussed in \S3, as \go\ increases, the ion-neutral coupling
increases and  the width of the shock
decreases.  This is evident in the scale of the $x$-axes in Figs.~\ref{gocomparisons}, \ref{shockconditions01}, 
\ref{shockconditions1}, \ref{shockconditions10}.  Likewise, as \go\ increases, 
roughly the same postshock temperatures are achieved with decreasing shock velocities.
Second, as \go\ increases, the increased photodissociation of molecules lowers the
peak postshock abundance of \water, OH, and \oxy, thus requiring greater \mh\ column 
densities to produce the same optically-thin \oxy\ line flux (see Table~2).

Within the observational uncertainties, all three models in Table~2 can explain both the measured \oxy\
line fluxes and line ratios; is one model preferred?  One possible discriminator is 
the degree to which each model agrees with 
estimates of the total \mh\ column density and number density toward Peak 1.
\citet{Rosenthal00} estimate the \mh\ column density to be $\approx\:$2\ti 10$^{22}$~\cms\ based on the
observation of 56 \mh\ ro-vibrational and pure rotational lines toward Peak 1 using the 
{\em Infrared Space Observatory}'s Short Wavelength Spectrometer \citep{deGraauw96}.
However, because the lowest-lying \mh\ pure rotational transition lies 510$\:$K above the ground state,
estimates based on \mh\ lines are more sensitive to the warm and hot gas and much less so to the cooler
(i.e., $T <\,$100$\:$K) gas where the \oxy\ emission peaks.

The cooler gas is better, if less directly, sampled by $^{12}$CO and its isotopologues.  
Using multi-line transitions of \nco, \ico, and \cio, \citet{Peng12} derive a \nco\ column density
of between 5\ti 10$^{18}$~\cms\ and 10$^{19}$~\cms\ toward Peak 1, in good agreement with
earlier estimates of \citet{Gonzalez02}.  This
corresponds to an \mh\ column density of between 6\ti 10$^{22}$~\cms\ and 
1.2\ti 10$^{23}$~\cms, assuming a \nco\ abundance of 8\ti 10$^{-5}$ \citep{Wilson92}.
However, even these studies rely on transitions with upper level temperatures above the ground state
$>\,$32$\:$K and may exclude contributions from a colder component.

Estimates of the preshock \mh\ number density vary depending on the tracer being modeled.
Generally, estimates exceed 10$^5$~\cmc, and can range as high as 10$^8$~\cmc\ 
\citep[cf.][]{Kaufman96a, Rosenthal00, Gonzalez02}.  None of the models we are aware of suggest
that preshock densities toward Peak 1 are less than 10$^4$~\cmc.

While none of the models we considered is in perfect agreement with the Peak 1 \mh\ column
density and number density estimates, the best-fit model in Table~2 that assumes \go$\,=\:$1 comes closest.
Moreover, the shock velocity of this model (i.e., 23\kms) is within the range of typical shock speeds
in OMC-1 of 30$\,\pm\,$10\kms\ \citep{Kristensen03, Kristensen07}.

The best-fit \go$\,=\:$0.1 model is a good approximation to the
fully-shielded $C$-type shock models that have traditionally been used to model a wide variety of
molecular emission toward outflows.  We show that fully-shielded (i.e., \go$\,\leq\,$0.1) models fail to 
reproduce the \oxy\ line intensities and ratios with shock conditions inferred from other data;
however, allowing for the presence of even modest amounts of FUV radiation (i.e., \go$\,\simgt\,$1) provides 
a better fit to the observations.

Finally, we note that the circumstances that explain the \oxy\ emission toward Orion \mh\ Peak 1
may also explain the rarity of similar detections in other sources.  First, the shocked gas emission around Peak 1
subtends $\sim\,$20\asec, larger than for most outflows.  Since the strength of the optically-thin line emission
within the beam scales with the \oxy\ column density\,\ti the area of the emitting region, the column density requirements
are eased if the emitting area is large, as toward Peak 1.  Second, the presence of strong \water\ 
maser emission toward the 
position of the \oxy\ detections is evidence not only for the presence of shocks, but for some shocks 
propagating in the plane of the sky.  High postshock temperatures promote the production of \oxy\
well above abundance levels predicted for cold quiescent gas, while the geometry (in the plane of the sky)
provides the observer with a potentially large line-of-sight column of gas with this relatively high \oxy\
abundance.  Of course, shocks of the sort described here would be expected to produce enhanced optically thin \oxy\ 
emission regardless of whether these shocks were propagating perpendicular to our line of sight or not.  However, if
such shocks produce \oxy\ integrated intensities comparable to what was detected toward Peak 1, but with propagation angles
closer to our line of sight, the \oxy\ line profiles would appear broader.  Thus, the same total line flux would be spread
over a greater number of velocity bins with a corresponding reduction in the line center amplitude.  
For weak emission, such as that exhibited by the observed \oxy\ lines, narrow lines are generally much easier 
to detect than lower amplitude broad lines, particularly in the presence of noise. 
The rarity of \oxy\ detections may therefore result from the need to simultaneously satisfy four conditions: (1)
an \oxy\ emitting area that fills a non-negligible fraction of a beam; (2) the presence of $C$-type shocks 
having peak postshock temperatures sufficient
to drive the efficient production of \water; (3) illumination by FUV radiation that can both enhance the preshock
gas-phase O-nuclei abundance and photodissociate postshock \water; and, (4) a shock propagation vector
close to perpendicular to the line of sight that naturally provides narrow line profiles.

\vspace{14mm}

We wish to acknowledge helpful discussions with Paul Goldsmith.  This work was supported by NASA through
Astrophysics Data Analysis Program grant No.~NNX13AF16G.

\clearpage

\bibliographystyle{plainnat}

\begin{thebibliography}{}\setlength{\itemsep}{1,8mm}

\vspace{1.7mm}

\bibitem[Beckwith \etal(1978)]{Beckwith78} Beckwith, S., Persson, S.E., Neugebauer, G., \& Becklin, E.E. 1978,
\apj, 223, 464

\bibitem[Bergin \etal(1998)]{Bergin98} Bergin, E. A., Melnick, G. J., \& Neufeld, D. A.
1998, \apj, 499, 777

\bibitem[Bergin \etal(2000)]{Bergin00} Bergin, E. A., Melnick, G. J., Stauffer, J. R. \etal\ 2000, \apj, 539, L129

\bibitem[Chen \etal(2014)]{Chen14} Chen, J-H., Goldsmith, P. F., Viti, S., \etal\ 2014, to appear in \apj

\bibitem[Cunningham(2006)]{Cunningham06} Cunningham, N. J. 2006, PhD Thesis, University of Colorado

\bibitem[de Graauw \etal(1996)]{deGraauw96} de Graauw Th., Haser L., Beintema D., \etal\ 1996, \aap, 315, L49


\bibitem[de Graauw \etal(2010)]{deGraauw10} de Graauw, Th., Helmich, F. P., Phillips, T. G., \etal\
 2010, \aap, 518, L6

\bibitem[Draine(1983)]{Draine83a} Draine, B.T. 1983, \apj, 270, 519

\bibitem[Draine \etal(1983)]{Draine83b} Draine, B.T., Roberge, W.G., \& Dalgarno, A. 1983, 
\apj, 264, 485

\bibitem[Flower \& Pineau des Flor\^{e}ts(1994)]{Flower94} Flower, D. R., \& Pineau des Flor\^{e}ts, G. 1994, \mnras,
268, 724

\bibitem[Flower \& Pineau des Flor\^{e}ts(2010)]{Flower10} Flower, D. R., \& Pineau des Flor\^{e}ts, G. 2010, \mnras,
406, 1745

\bibitem[Gautier \etal(1976)]{Gautier76} Gautier, T.N., III, Fink, U., Treffers, R.P., \& Larson, H.P. 1976, \apjl, 207, L129

\bibitem[Genzel \etal(1981)]{Genzel81} Genzel, R., Reid, M. J., Moran, J. M., \& Downes, D. 1981, \apjs, 30, 145


\bibitem[Goldsmith \& Langer(1978)]{Goldsmith78} Goldsmith, P. F., \& Langer, W. D. 1978, \apj, 222, 881


\bibitem[Goldsmith \etal(2002)]{Goldsmith02} Goldsmith, P. F., Li, D., Bergin, E. A., \etal\ 2002, \apj, 576, 814

\bibitem[Goldsmith \etal(2011)]{Goldsmith11} Goldsmith, P. F., Liseau, R., Bell, T. A., \etal\ 2011, \apj, 737,
96

\bibitem[Goldsmith \etal(2000)]{Goldsmith00} Goldsmith, P. F., Melnick, G.J., Bergin, E.A., \etal\ 2000, \apjl, 539, L123


\bibitem[Goldsmith \etal(1985)]{Goldsmith85} Goldsmith, P. F., Snell, R. L., Erickson, N. R., \etal\ 1985, \apj,
289, 613

\bibitem[Gonz\'{a}lez-Alfonso \etal(2002)]{Gonzalez02} Gonz\'{a}lez-Alfonso, E., Wright, C. M., Cernicharo, J.,
\etal\  2002, \aap, 386, 1074


\bibitem[Habing(1968)]{Habing68} Habing, H. J. 1968, {\em Bull. Astron. Inst. Netherlands}, 19, 421

\bibitem[Herrmann \etal(1997)]{Herrmann97} Herrmann, F., Madden, S. D., Nikola, T., \etal\ 1997, \apj, 481, 343


\bibitem[Hollenbach \etal(2009)]{Hollenbach09} Hollenbach, D.J., Kaufman, M. J., Bergin, E.A., \& 
Melnick, G.J. 2009, \apj, 690, 1497


\bibitem[Jim\'{e}nez-Serra \etal(2008)]{Jimenez-Serra08} Jim\'{e}nez-Serra, I., Caselli, P., Mart\'{i}n-Pintado, M., \etal\ 2008,
\aap, 482, 549

\bibitem[Kaufman \& Neufeld(1996a)]{Kaufman96a} Kaufman, M. J. \& Neufeld, D. A. 1996a, \apj, 456, 611

\bibitem[Kaufman \& Neufeld(1996b)]{Kaufman96b} Kaufman, M. J. \& Neufeld, D. A. 1996b, \apj, 456, 250


\bibitem[Kristensen \etal(2003)]{Kristensen03} Kristensen, L. E., Gustafsson, M., Field, D., \etal\ 2003, 
\aap, 412, 727

\bibitem[Kristensen \etal(2007)]{Kristensen07} Kristensen, L. E., Ravkilde, T. L., Field, D., 
\etal\ 2007, \aap, 469, 561

\bibitem[Kristensen \etal(2012)]{Kristensen12} Kristensen, L. E., van Dishoeck, E. F., Bergin, E. A., 
\etal\ 2012, \aap, 542, A8

\bibitem[Larsson \etal(2007)]{Larsson07} Larsson B., Liseau R., Pagani L., et al. 2007, \aap, 466, 999

\bibitem[Lique(2010)]{Lique10} Lique, F. 2010, {\em J. Chem. Phys.}, 132, 044311

\bibitem[Liseau \etal(2012)]{Liseau12} Liseau, R., Goldsmith, P. F., Larsson, B., \etal\ 2012, \aap, 541,
A73

\bibitem[Marconi \etal(1998)]{Marconi98} Marconi, A., Testi, L., Natta, A., \& Walmsley, C. M. 1998, \aap, 330, 696

\bibitem[McElroy \etal(2013)]{McElroy13} McElroy, D., Walsh, C., Markwick, A.~J., \etal\ 2013, \aap, 550, A36

\bibitem[Melnick \etal(2000)]{Melnick00} Melnick, G. J., Stauffer, J. R., Ashby, M. L. N., \etal\ 2000, \apj, 539, L77

\bibitem[Melnick \etal(2012)]{Melnick12} Melnick, G. J., Tolls, V., Goldsmith, P. F., \etal\ 2012, \apj, 752, 26

\bibitem[Neufeld \etal(2014)]{Neufeld14} Neufeld, D. A., Gusdorf, A., G\"{u}sten, R., \etal\ 2014, \apj, 781, 102


\bibitem[Neufeld, Lepp \& Melnick(1995)]{Neufeld95} Neufeld, D.A., Lepp, S. \& Melnick, G. J. 1995, \apjs,
100, 132

\bibitem[Neufeld \& Melnick(1991)]{Neufeld91} Neufeld, D. A., \& Melnick, G. J. 1991, \apj, 368, 215

\bibitem[Neufeld \etal(2013)]{Neufeld13} Neufeld, D. A., Wu, Y., Kraus, A., \etal\ 2013, \apj, 769, 48


\bibitem[Nordh \etal(2003)]{Nordh03} Nordh, H. L.,  von Sch\'{e}ele, F., Frisk, U., \etal\ 2003, \aap, 402, L21

\bibitem[Pagani \etal(2003)]{Pagani03} Pagani L., Olofsson A. O. H., Bergman P., et al. 2003, \aap, 402, L 77

\bibitem[Peng \etal(2012)]{Peng12} Peng, T.-C., Wyrowski, F., Zapata, L. A., G\"{u}sten, R., \& Menten, K. M. 
2012, \aap, 538, A12

\bibitem[Pilbratt \etal(2010)]{Pilbratt10} Pilbratt, G. L., Riedinger, J. R., Passvogel, T., et al. 2010, \aap, 518, L1

\bibitem[Roelfsema \etal(2012)]{Roelfsema12} Roelfsema, P. R., Helmich, F. P., Teyssier, D., et al. 2012, \aap,
537, A17

\bibitem[Rosenthal, Bertoldi \& Drapatz(2000)]{Rosenthal00} Rosenthal, D., Bertoldi, F., \& Drapatz, S. 2000,
 \aap, 356, 705

\bibitem[Sandqvist \etal(2008)]{Sandqvist08} Sandqvist A., Larsson B., Hjalmarson, \AA., et al. 2008, \aap, 
482, 849

\bibitem[van Kempen \etal(2010)]{vanKempen10} van Kempen, T. A., Kristensen, L.E., Herczeg, G.J., \etal\ 2010, \aap, 518, 121



\bibitem[Walmsley \etal(2000)]{Walmsley00} Walmsley, C. M., Natta, A., Oliva, E., \& Testi, L. 2000, \aap, 364, 301

\bibitem[Wilson \& Matteucci(1992)]{Wilson92} Wilson, T. L., \& Matteucci, F. 1992, {\em A\&ARv}, 4, 1

\end{thebibliography}

\clearpage

\phantom{0}

\vspace{0.6in}

\begin{table}[htdp]
\begin{center}
TABLE~1.~~Summary of \oxy\ Observations Toward Orion \mh\ Peak 1\\*[1.4mm]
\begin{tabular}{ccccccc} \hline \\*[-4.3mm] \hline
\rule{0mm}{6mm}Rest  &   &  Energy Above  &  \phantom{$^{\, a}$}Integrated$^{\, a}$  &  
\phantom{$^{\, a}$}Line$^{\, a}$  & LSR  &  \phantom{$^{\, b}$}HIFI Beam$^{\, b}$ \\
~Frequency~  & Transition  &  Ground State  &  Intensity  &  Width  &  Velocity  & ~Diameter~ \\*[0.5mm]
(GHz)  &  ($N, J \rightarrow N, J$)  &  ($E_u/k$)  &  ~(K km s$^{-1}$)~  &  (km s$^{-1}$)  &  
(km s$^{-1}$)  &  (arcsec) \\*[1.5mm] \hline
\rule{0mm}{6mm}487.249  &  3, 3 \dash\ 1,2  &  26$\:$K  &  0.081  &  3.1  &  10.2  &  44.7  \\*[1.5mm]
773.840  &  5, 4 \dash\ 3, 4  &  61$\:$K  &  0.154  &  1.7  &  11.0  &  28.2  \\*[1.5mm]
1120.715~~  &  7, 6 \dash\ 5, 6  &  115$\:$K~~  &  0.043  &  0.8  &  11.0  &  18.9  \\*[1.5mm] \hline 
\multicolumn{6}{l}{\rule{0mm}{5.8mm}$^a\:$After \citet{Chen14}; $\:^b\:$See \citet{Roelfsema12}.}
\end{tabular}
\end{center}
\label{default}
\end{table}%

\clearpage

\begin{sidewaystable}[htdp]
\begin{center}
TABLE~2.~~Best-Fit Shock Models\\*[1.6mm]
\begin{tabular}{lcccccccc} \hline \\*[-4.3mm] \hline
\rule{0mm}{7mm}  &  Log$_{\rm 10}$ Preshock  &  Shock  &  Peak Postshock  &   
\multicolumn{5}{c}{\raisebox{1.1mm}{\underline{\phantom{0}Conditions at Peak of the 487, 774, and 1121 GHz Emission\phantom{0}}}} \\
~~\go\hspace{7mm}  &  \mh\ Density  & ~~Velocity~~  &   ~~O$_2$ Abundance~~ & Transition & \nh\ &  
$N$(\mh)  &  $\chi$(\oxy)  & Temp. \\
  &  (cm$^{-3}$)  &  (km s$^{-1}$)  &  (rel. to \mh)   &  (GHz)  &  (cm$^{-3}$)  &  (cm$^{-2}$)  & (rel. to \mh)  &  
  (K)  \\*[1.5mm] \hline
\rule{0mm}{6mm}~\phantom{1}0.1 \dotfill\  &  3.4  &  28  &  6.0\ti 10$^{-6}$  &   &  &  &  &  \\
  &  &  &  & 487  & 5.0 \ti 10$^{4}$  &  2.1 \ti 10$^{23} \cdot \gamma$   &  6.0 \ti 10$^{-6}$  &  24 \\*[1mm]
  &  &  &  & 774  & 4.9 \ti 10$^{4}$  &  2.1\ti 10$^{23} \cdot \gamma$   &  5.1 \ti 10$^{-6}$  &  34 \\*[1mm]
  &  &  &  & 1121~~  & 4.7 \ti 10$^{4}$  &  2.0 \ti 10$^{23} \cdot \gamma$   &  3.3 \ti 10$^{-6}$  &  51 \\*[2.5mm]
~\phantom{0}1 \dotfill\  &  4.9  &  23  &  1.7\ti 10$^{-6}$  &   &  &  &  &  \\
  &  &  &  & 487  & 1.3 \ti 10$^{6}$  &  3.7 \ti 10$^{23} \cdot \gamma$   &  1.7 \ti 10$^{-6}$  &  24 \\*[1mm]
  &  &  &  & 774  & 1.3 \ti 10$^{6}$  &  3.6\ti 10$^{23} \cdot \gamma$   &  1.4 \ti 10$^{-6}$  &  39 \\*[1mm]
  &  &  &  & 1121~~  & 1.2 \ti 10$^{6}$  &  3.5 \ti 10$^{23} \cdot \gamma$   &  1.0 \ti 10$^{-6}$  &  61 \\*[2.5mm]

~10 \dotfill\  &  6.0  &  14  &  1.0\ti 10$^{-6}$  \\
  &  &  &  & 487  & 9.5 \ti 10$^{6}$  &  6.2 \ti 10$^{23} \cdot \gamma$   &  1.0 \ti 10$^{-6}$  &  26 \\*[1mm]
  
  &  &  &  & 774  & 9.3 \ti 10$^{6}$  &  6.0 \ti 10$^{23} \cdot \gamma$   &  9.3 \ti 10$^{-7}$  &  47 \\*[1mm]
    
  &  &  &  & 1121~~  & 9.0 \ti 10$^{6}$  &  5.8 \ti 10$^{23} \cdot \gamma$   &  7.7 \ti 10$^{-7}$  &  73 \\*[1mm] \hline
\end{tabular}
\end{center}
\label{default}
\end{sidewaystable}%

\clearpage

\clearpage

\clearpage

\clearpage

\clearpage

\clearpage

\clearpage

\clearpage

\clearpage

\clearpage

\clearpage

\end{document}